\documentclass[conference]{IEEEtran}
\IEEEoverridecommandlockouts
\usepackage{cite}
\usepackage{amsmath,amssymb,amsfonts}
\usepackage{algorithmic}
\usepackage{graphicx}
\usepackage{textcomp}
\usepackage{xcolor}
\usepackage{subfigure}
\usepackage{array}
\usepackage{booktabs}
\usepackage{multirow}
\usepackage{multicol}
\usepackage{threeparttable}
\def\BibTeX{{\rm B\kern-.05em{\sc i\kern-.025em b}\kern-.08em
    T\kern-.1667em\lower.7ex\hbox{E}\kern-.125emX}}
\begin{document}

\title{Staggering and Fragmentation for Improved Large Message Handling in libp2p GossipSub\\
}

\author{\IEEEauthorblockN{1\textsuperscript{st} Muhammad Umar Farooq}
\IEEEauthorblockA{\textit{Vac Research} \\
\textit{Institute of Free Technology (IFT),}\\
Singapore \\
farooq@status.im}
\and
\IEEEauthorblockN{2\textsuperscript{nd} Tanguy Cizain}
\IEEEauthorblockA{\textit{Vac Research} \\
\textit{Institute of Free Technology (IFT),}\\
Singapore \\
tanguy@status.im}
\and
\IEEEauthorblockN{3\textsuperscript{rd} Daniel Kaiser}
\IEEEauthorblockA{\textit{Vac Research} \\
\textit{Institute of Free Technology (IFT),}\\
Singapore \\
danielkaiser@status.im}
}

\maketitle

\begin{abstract}
The libp2p GossipSub protocol leverages a full-message mesh with a lower node degree and a more densely connected metadata-only (gossip) mesh. This combination allows an efficient dissemination of messages in unstructured peer-to-peer (P2P) networks. However, GossipSub needs to consider message size, which is crucial for the efficient operation of many applications, such as handling large Ethereum blocks. This paper proposes modifications to improve GossipSub's performance when transmitting large messages.
We evaluate the proposed improvements using the shadow simulator.
Our results show that the proposed improvements significantly enhance GossipSub's performance for large message transmissions in sizeable networks.      
\end{abstract}


\section{Introduction}
Current networked application architectures are driven by the need to share information, where information is usually available at logically centralized locations (servers). The focus is on coupling these locations with end users (clients). Clients can always look for their intended servers by relying on a less dynamic host-centric communication model. This model requires application-specific massive infrastructure deployment and enforces many limitations \cite{xylomenos2013survey}. However, modern decentralized application paradigms like Web3, Blockchain, and information-centric networks (ICN) drive the networked application architectures towards decoupled client-server interactions. Unlike their predecessors, such applications place content (and relative functionality for accessing and modifying this content) at the center. As a result, the share of Internet usage for information dissemination among distributed systems is rapidly increasing compared to traditional client-server interactions. Many distributed networking applications have emerged during the last decade \cite{benet2014ipfs, status2023app}. Such applications can seamlessly connect resources and users in a scalable manner. This is because participating peers also bring in computational and network resources and do not rely on any central arrangements for successful operation, hence termed P2P systems. The idea is to use these resources for valuable tasks. For this purpose, P2P applications form direct communication links between the communicating peers. The selection of links can be random or inferred through some peer discovery mechanism, and this selection has a noticeable impact on the protocol's performance. For instance, peering among nodes within the same ISP and peering among nodes over cross-oceanic links can significantly impact the latency of the network. These peering links are transport layer connections between hosts, thus representing an overlay network. The overlay network has self-forming and self-healing capabilities, and the peering algorithm governs the construction and maintenance of the overlay network. The peering algorithm is not trivial and requires participating nodes to collaborate in a trustless environment to reach a consensus. Generally, P2P networks can be classified into structured and unstructured P2P networks.

Structured P2P networks arrange peers in an organized overlay. These networks typically use distributed hash tables (DHT) or similar mechanisms to facilitate seamless P2P interactions. While systems like Skype previously used supernodes to fulfill additional roles, modern systems more commonly depend on DHT and other decentralized methods to manage peer connections with less reliance on special-purpose nodes.

Unstructured (or pure) P2P networks operate without structural overlay support, and peers join the network without additional attributes. Random connections between peers carry out resource discovery and information dissemination. Peering algorithms direct these connections and their lifetimes. These algorithms vary from flooding variants \cite{lin1999gossip, benet2016floodsub} to knowledge-based (or probabilistic) topologies \cite{escalante2005rng, tumas2022probabilistic}. However, publish-subscribe (PubSub) \cite{eugster2003many, git2020pubsub} methods have proven more useful. In PubSub, nodes advertise attributes of data they will publish (topics) to the network. Based on this information, peers subscribe to their topics of interest and information from sources (publishers) is disseminated to the subscribers through overlay network(s). A large number of PubSub variants are available. In Floodsub \cite{benet2016floodsub}, on receiving a message related to a specific interest group, a peer sends a copy to all known peers subscribed to that topic. Information travels through all possible paths, forming a fully connected overlay mesh. This approach provides minimum message dissemination latency and maximum resilience against network attacks. However, extensive redundant transmissions can lead to severe network congestion. GossipSub \cite{vyzovitis2020gossipsub} introduces an efficient solution for the information propagation problem in unstructured P2P networks. In GossipSub, every node peers with $D$ other nodes to form a full-message mesh and $K$ other nodes to create a gossip mesh, where $K \geq D$. All messages flow through the full-message mesh, and metadata flows through the gossip mesh. The metadata contains IHAVE messages, announcing IDs of seen messages. Nodes can fetch any unseen messages using IWANT requests. In conjunction with gossip mesh, the full-message mesh assures resilience against attacks and a message dissemination latency comparable to FloodSub. However, redundant large message transmissions compromise GossipSub's performance by substantially increasing message dissemination latency and bandwidth utilization. To meet these challenges, modifications are proposed to the GossipSub protocol to enhance its performance for large messages. It is worth mentioning that the proposed changes will only take effect if the message size increases a certain threshold. The major contributions of this article are as follows:

\begin{enumerate}
    \item Message transmissions follow a store-and-forward process at all peers, which is inefficient in the case of large messages. We parallelize message transmissions by partitioning large messages into smaller fragments, letting intermediate peers relay these fragments as soon as they receive them.

    \item We find that simultaneously relaying a message to all peers can increase store-and-forward delay. Therefore, we propose a message-staggering strategy that accelerates message transfers to individual peers.
\end{enumerate}    

Moreover, we implement the IDONTWANT message proposal \cite{Nashatyrev2023idontwant} to curtail redundant transmissions of large messages. This works by notifying other peers that we have already received the message. The combined use of staggering, fragmentation, and IDONTWANT messages results in significant performance improvements. The rest of the article is organized as follows: in section \ref{S1}, the current state of the art is provided. In section \ref{S2}, proposed GossipSub modifications are suggested. Performance evaluation results are presented in section \ref{S3}, and section \ref{S4} concludes the article.

\section{Related Work} \label{S1}
To put the proposed improvements in proper context, we briefly review different strategies for disseminating large volumes of information in P2P networks. We consider overlay designs, peering improvements, message coding, pull-based operations, etc. The focus is to find solutions that yield lower latency and minimize bandwidth utilization while assuring GossipSub-like resilience. 

Several studies aim to improve performance by effectively handling high traffic volumes. GoCast \cite{tang2005gocast} uses an overlay mesh with a multicast tree embedded in it. The messages propagate through the tree links. To avoid failures, peers also randomly gossip across the overlay links. The peers can request any missing messages through the overlay link. In \cite{savolainen2020streamr}, authors propose a decentralized PubSub network that forms a separate overlay for each topic (called stream). The overlay is built on top of the Ethereum infrastructure layer. Brokers manage the main network operation. Peers subscribe with brokers for their streams of interest. Sharding is used to partition large events, and these partitions are stored at different brokers to achieve better load balancing. Distributed Publish \& Subscribe for IoT (DPS) \cite{intelDPS4IoT2019} is an MQTT-inspired PubSub architecture for fully distributed P2P networks. However, unlike MQTT, the role of a broker is short-lived. It is maintained for sending a single subscription/message while using hop-by-hop message relaying, enabling subscribers to receive their intended information. Epidemic PubSub (libp2p-episub) \cite{vyzo2018episub} minimizes bandwidth and computational resource wastage by curtailing redundant transmissions. The algorithm maintains a small active view and a large passive view. The active view nodes are periodically swapped with better nodes from the passive view. Message transmissions are carried out using epidemic broadcast trees \cite{leitao2007epidemic}, where all peers are initially placed in the eager push set and then periodically moved to the lazy push set. In \cite{de2019systems}, authors use a similar approach and create an overlay with a small clustering coefficient using the X-BOT algorithm \cite{leitao2012x}. Pemcast \cite{tumas2022probabilistic} minimizes redundant transmissions while assuring resilience and fair-length paths. Each pemcast node maintains an r-hop view of the neighborhood and recognizes nodes at r-hop distance as edge nodes. A fanout number of edge nodes is selected for message dissemination, and a multicast sub-tree is formed to cover the fanout nodes. These forwarders further relay this message through their edge (fanout) nodes. The acknowledgment from the receiver helps confirm the appropriate paths. In \cite{frey2022differentiated}, authors prose a two-phase epidemic broadcast protocol to minimize latency and attain differentiated consistency. The protocol probabilistically divides the peers into primary and secondary peers. This gossip primary and secondary (GPS) algorithm first assures fast convergence between primary peers, and then a moderate convergence rate is adapted for secondary peers. The authors successfully evaluated the proposed scheme on large P2P networks and the Ethereum blockchain. In \cite{zaarour2022openpubsub}, authors realize the scale of content space between publishers and subscribers in P2P networks. The authors propose a content-based approximate semantic PubSub model that supports hybrid event routing. This is achieved by using rendezvous routing, gossiping, and clustering to reduce message overhead and redundant transmissions. FRING \cite{qiu2022geography} forms a fractal ring structure by placing nodes into rings on the basis of geographical proximity. Multiple nodes from each ring act as representatives and are recursively connected with the upper ring(s) nodes. Message propagation involves spreading messages to random representatives in the ring that initiate intra-ring broadcasts, and message spreading in the upper ring. A broadcast-down process assures network-wide reachability once the message reaches the top level. PeerDAS \cite{Danny2024eip7594} employs data availability sampling to ensure that essential blob data remains accessible. It allows downloading only a subset of information using additional discovery and request features integrated into the GossipSub (still a work in progress) \cite{pop2024pr617}. 

The above works mainly focus on enhancing scalability by reducing the degree of the overlay network or by limiting information dissemination. This is achieved through rendezvous routing, broker placement, clustering, pull-based operation, etc. Mechanisms like smaller number of $f_{out}$ (outgoing) links, source peer-set randomization, and probabilistic/weighted peer matching are also suggested in \cite{shaleva2021efficient, agostinho2022smartpubsub}. Network coding approaches \cite{bromberg2019multisource, haeupler2011analyzing, yu2007massive} benefit from the redundant transmissions carried out by the gossip protocols to send linear combinations of several messages in place of unique plain messages. The work in \cite{sanghavi2007gossiping} further highlights the performance gain achieved by partitioning and disseminating a large file in the network, compared to a complete file transfer.

Minimizing the degree of the full-message mesh can reduce bandwidth utilization by curtailing the number of redundant transmissions. However, these mechanisms have certain shortcomings. For instance, reducing the degree of the overlay network may increase the number of messages taking longer than usual paths. Similarly, a small node degree exposes the network to many attacks. Besides, having a rendezvous point or clustering is ineffective in unstructured P2P networks. Therefore, algorithms that minimize redundant traffic while maintaining the same node degree level are highly desirable. At the same time, such algorithms empower peers to adapt their message-forwarding behavior according to message sizes and frequency. This adaptability is highly desirable for unstructured P2P networks to be equally effective against varying network characteristics and ever-changing application needs.

\section{Proposed Modifications} \label{S2}
GossipSub uses redundant transmissions to ensure resilience against adversaries, but this can overwhelm outgoing message queues when handling large messages. To address this issue, we propose the following improvements to the protocol operation. It is important to note that these changes do not compromise the GossipSub resilience and take effect only when the message size exceeds a specified threshold.

\begin{figure*}[t]
    \centering
    \subfigure[Message propagation from A]{
        \includegraphics[width=0.3\linewidth]{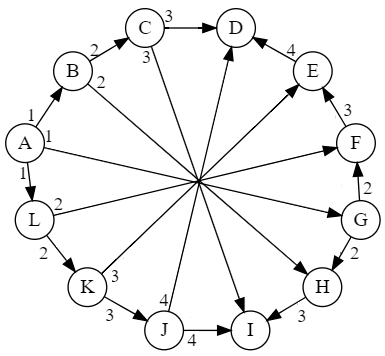}
    }
    \subfigure[Order of message delivery at peers]{
        \includegraphics[width=0.3\textwidth]{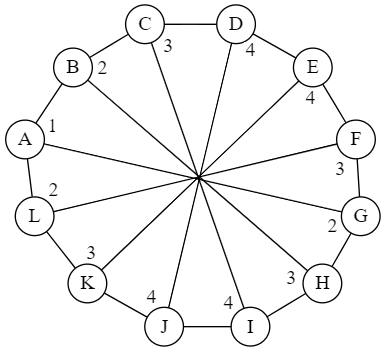}
    }
    \subfigure[Duplicate full-message mesh transmissions]{
        \includegraphics[width=0.3\textwidth]{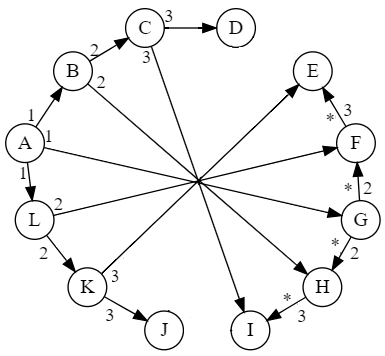}
    }
    \subfigure[Effect of IDONTWANT message]{
        \includegraphics[width=0.3\textwidth]{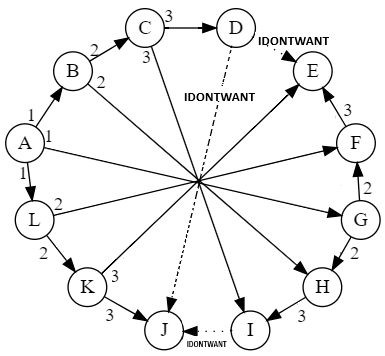}
    }
    \subfigure[Staggered message propagation from A]{
        \includegraphics[width=0.3\linewidth]{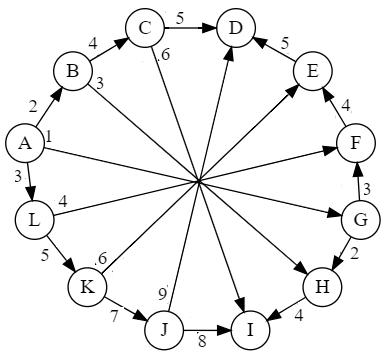}
    }
    \subfigure[Message-staggering with IDONTWANTs]{
        \includegraphics[width=0.3\linewidth]{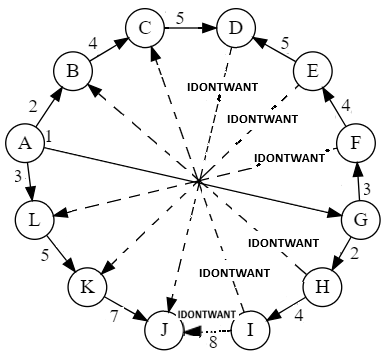}
    }    
    \caption{GossipSub full-message mesh with D=3}
    \label{fig:gossipsub}
\end{figure*}

\subsection{Message Fragmentation} 
Ignoring any message processing or similar delays, we can approximate network-wide message dissemination time as $\tau \approx (\tau_p + \tau_{tx}) \times H$, where $\tau_p$ is average link latency and $\tau_{tx}$ is transmission time, denoted as $\tau_{tx} = \frac{S}{R}$ with $S$, $R$ being data size and data rate respectively. $H$ represents network diameter (number of hops required to reach the farthest node). At the same time, each node performs roughly $D$ transmits/receives for every published message. As a result, message transmission time $\tau_{tx}$ increases noticeably with the message size. Message fragmentation allows the partitioning of a large message into smaller fragments. As a result, single fragment transmission time drops to $\frac{\tau_{tx}}{n}$, where $n$ represents the number of fragments. Since message relaying requires a store-and-forward process, sending a message in the form of fragments allows immediate relaying of received fragments by the intermediate nodes while the sender is still transmitting the remaining fragments. This approach reduces message transmission time across $H$-hops (network diameter) from $\tau_{tx} \times H$ to $\tau_{tx} \times \frac{2H-1}{n}$, which demonstrates a significant decrease in the store-and-forward delay associated with large message forwarding.

It is worth mentioning that some applications, like Ethereum, require each fragment to be individually verifiable. This constraint adds additional processing time at each hop and necessitates altering the protection mechanisms. Therefore, message fragmentation for such applications requires careful tradeoff analysis between time and risks, or alternatively, fragmentation at the application level can be considered.

\subsection{IDONTWANT Message to Mitigate Duplicates} \label{chokemsg}
We implement the IDONTWANT\footnote{IDONTWANT and IHAVE messages are similar. However, peers share IHAVE announcements during heartbeat intervals within their gossip mesh. By sending an IDONTWANT, a peer immediately informs its full-message mesh members that it has successfully received the announced message.} message proposal \cite{Nashatyrev2023idontwant} to prevent mesh members from resending large messages to peers who have already received them. On receiving a message larger than the specified threshold, every peer sends an IDONTWANT announcement to the remaining mesh members, asking them to refrain from resending the same message. 

The use of IDONTWANT messages requires specific considerations. For instance, Fig.\ref{fig:gossipsub}-(a) presents message propagation from peer A through full-message mesh with node degree three. The message spreads throughout the network in four iterations (rounds). The labels on the directed edges indicate the round during which each transmission occurs. The labels on the vertices in Fig.\ref{fig:gossipsub}-(b) show the round where that peer receives the first copy of the message. Assuming all edges introduce similar latency and peers simultaneously retransmit the message to their mesh members in each round\footnote{An IDONTWANT announcement can eliminate a duplicate transmission only if the sender receives and processes it before transmitting the corresponding message. We simplify this constraint by assuming that transmissions occur simultaneously in each round and IDONTWANT messages are sent immediately before the start of the next round.}, IDONTWANT messages can only prevent redundant transmissions scheduled for the next iteration. Fig.\ref{fig:gossipsub}-(c) illustrates this scenario, where node F receives the same message from nodes L and G in the second iteration. Therefore, sending an IDONTWANT message is not helpful. Such redundant edges are labeled with an asterisk (*). However, node D receives the first copy of the message from node C in the third iteration. Node D expects the same message from nodes E and J in the fourth iteration. Similarly, node I is scheduled to receive the same message from node J in the fourth iteration. IDONTWANT announcements can prohibit nodes E and J from resending the same message. The transmission of IDONTWANT messages from nodes D and I is depicted in Fig.\ref{fig:gossipsub}-(d). IDONTWANT messages can eliminate many redundant transmissions, especially during the later stages of message propagation. However, this mechanism is not helpful in the case of simultaneous receptions, as evidenced by Fig.\ref{fig:gossipsub}-(c). It is important to note that a longer transmission time can increase the likelihood of simultaneous message reception. Therefore, using IDONTWANT messages with message-staggering (detailed in section \ref{staggered}) can lead to significant performance improvement.

\subsection{Message-Staggering: Sequential Relaying of Messages} \label{staggered}
GossipSub nodes concurrently maintain transport layer connections with many peers. Relaying a message involves simultaneous resending to all successors, i.e., the full-message mesh members (excluding the ones from which we received the message or a corresponding IDONTWANT announcement). These messages are not immediately transferred. They are accepted and scheduled for transfer by the transport layer. After that, it becomes challenging for the application to abort the transmission of these messages. Therefore, receiving an IDONTWANT message at this stage does not affect the regular GossipSub operation (already scheduled redundant transmissions cannot be canceled). A solution to mitigate this problem is to stagger the order of message transmissions. In message-staggering, a message is sequentially forwarded to each successor. This allows more time for IDONTWANT announcements to be received from the remaining successors.

Message-staggering can also speed up message transfers to individual peers, which helps minimize store-and-forward delays. For instance, considering the same bandwidth, the time $\tau_D$ required for sending a message to D peers stays the same, even if we relay to all peers in parallel or send sequentially to the peers, i.e., $\tau_D = \sum_{i=1}^{D} \tau_i$. However, sequential relaying results in quicker message reception at individual peers ($\tau_1 \approx \frac{\tau_D}{D}$) due to bandwidth concentration for a single peer. So, the receiver can start relaying early to its successors while the original sender is still sending the message to other peers. As a result, after every $\frac{\tau_D}{D}$ milliseconds, the number of peers receiving the message increases by $2^X\ \forall\ X \in \{0, D-1\}$ and by $\sum_{k=X-D}^{X-1} \lambda_k\ \forall\ X \geq D$. Here, $X$ represents the message transmission round, and $\lambda_k$ represents the number of peers that received the message in round $k$. It is important to note that a realistic network imposes certain constraints on message-staggering. For instance, in a network with dissimilar peer capabilities, placing a slow peer (also in cases where many senders simultaneously select the same receiver) at the head of the transmission queue may result in head-of-line blocking. Higher link latency can also delay the propagation of large messages in staggered sending by increasing message reception time at individual peers. Similarly, a malicious peer can deliberately slow down the sending or receiving of messages. Therefore, instead of sequential sending, relaying to $K < D$ successors in parallel can help mitigate this problem.

In this article, we experiment with message-staggering for large messages only, as redundant large message transmissions can noticeably overwhelm the network. The use of IDONTWANT announcements with message-staggering can mitigate this problem. However, in practice, message-staggering can also be applied to smaller messages. Fig. \ref{fig:gossipsub}-(e) depicts the staggered relaying of a message to all subscribed peers. During each round, a single message is generated by each covered peer (peers that have already received the message). This transmission continues until all the covered peers have transmitted the message to their successors. The directed edges indicate message transmissions, and the labels on the edges indicate the round in which these transmissions are carried out. Fig. \ref{fig:gossipsub}-(f) indicates the issuance of IDONTWANT messages in conjunction with staggered sending. By issuing an IDONTWANT message, a node can inform its peers to preserve their uplink bandwidth by eliminating several redundant transmissions. Eventually, the message propagation completes in less time. However, using staggered sending with IDONTWANT messages does not completely eliminate redundant transmissions.

\subsection{Considering the Impact of TCP Congestion Avoidance}
TCP's message transmission mechanisms share the same fundamental principles of reliable data delivery, whereas congestion avoidance algorithms are different across various implementations of TCP. Selecting an appropriate congestion avoidance algorithm is vital in deciding achievable data rate and latency during large message transmissions. However, this choice typically depends on the computing environment and has system-wide implications. Modern computing environments usually use TCP cubic, compound TCP, datacenter TCP, or similar variants. These variants initiate connections with a small congestion window ($C_{wnd}$), which rises with the data flow. Consequently, sending large messages through floodpublish or IWANT replies may take longer due to a smaller $C_{wnd}$ at less frequent links. A smaller $C_{wnd}$ can also dilute the benefits of message-staggering by slowing down message transfers to individual peers. Moreover, some TCP variants may reset their $C_{wnd}$ if a link remains idle for an extended period \cite{chu2013rfc}. In some applications like Ethereum, we periodically use a link to its full potential and leave it idle (or at a slow packet rate) for the rest of the time. In the case of prolonged inactivity, TCP may restart $C_{wnd}$ probing. Libp2p ping \cite{git2022libp2pping} provides a straightforward remedy to this problem. Pinging an inactive connection every few round-trip times ensures that the link stays warm. Similarly, in some environments, parameters like tcp\_slow\_start\_after\_idle may also help adjust TCP's behavior after inactive periods.

\begin{table}[t]
  \caption{Simulation Scenarios}
  \centering
  \begin{tabular}
  {|@{\hskip 0pt}>{\centering\arraybackslash}m{1.55cm}@{\hskip 0pt}|
  @{\hskip 0pt}>{\centering\arraybackslash}m{1.45cm}@{\hskip 0pt}|
  @{\hskip 0pt}>{\centering\arraybackslash}m{1.4cm}@{\hskip 0pt}|
  @{\hskip 0pt}>{\centering\arraybackslash}m{1.5cm}@{\hskip 0pt}|
  @{\hskip 0pt}>{\centering\arraybackslash}m{1.4cm}@{\hskip 0pt}|}
    \hline
    \textbf{Experiments}& \textbf{No. of Nodes}& \textbf{No. of Publishers}& \textbf{Message Size (KB)}& \textbf{Inter-Message Delay}\\
    \hline
    Scenario 1 &2000, 4000, ...., 12000& 12 & 200 & 3 sec\\
    \hline
    Scenario 2 & 1000 & 12 & 200, 400, ...., 1000 & 4 sec\\
    \hline
    Scenario 3 & 1000 & 22, 42, ...., 102 & 50 & 100 ms\\
    \hline
  \end{tabular}
  \label{tab:Scenarios}
\end{table}

\begin{table}[t]
  \caption{Simulation Parameters}
  \centering
  \begin{tabular}{|c|c|c|c|}
    \hline
    \textbf{Parameter}& \textbf{Value}&\textbf{Parameter}&\textbf{Value}\\
    \hline
    $D$ & 8 & $D_{low}$ & 6\\
    \hline
    $D_{lazy}$ & 6 & $D_{high}$ & 12\\
    \hline
    $D_{out}$ & $\frac{D_{low}}{2}$ & gossipFactor & 0.05 \\
    \hline
    Heartbeat Interval & 1000 ms & FloodPublish & false\\    
    \hline
    Stagger Interval & 200 ms & Muxer & yamux\\
    \hline
  \end{tabular}
  \label{tab:parameters}
\end{table}

\section{Results and Discussions} \label{S3}
The proposed modifications are implemented in nim-libp2p \cite{status2023nim-libp2p, stagger-nim-libp2p2023} and evaluated for performance in handling large messages. Message fragmentation is achieved by partitioning messages at the application level \cite{nimlibp2pShadowTestnode}. We use expanding coverage latency, latency deviation, and network-wide bandwidth utilization as performance evaluation metrics. Expanding coverage latency $L_{cov}^i$ represents the time required to reach $i$ nodes, where $i \in \{15\%, 85\%, 100\%\}$. Network-wide message dissemination time $L_{cov}^{100}$ measures the time it takes for a message to reach the entire network. In the remaining part of this article, we also refer to $L_{cov}^{100}$ as latency or $L_{j}^N$, with $j$ indicating the size of the published message and $N$ indicating the size of the network. Network-wide bandwidth utilization $B_N$ provides the total traffic volume, including control traffic and actual data transmissions. 
Three simulation scenarios are considered: 1) the number of publishers and message size remain constant, and the network size gradually increases. 2) The number of publishers and the network size remain constant while the message size increases. 3) The network and message sizes remain unchanged while the number of publishers increases. In all simulation scenarios, every publisher publishes exactly one message to the network. The subsequent publisher waits for a predefined inter-message delay before sending the next message. Inter-message delay helps achieve different traffic patterns, and having multiple publishers contributes to fair performance evaluations, as every published message takes a different path. $L_{cov}^{100}$ averages latency estimates for published messages and $\delta_L$ provides latency deviation. Detailed simulation scenarios are presented in Table \ref{tab:Scenarios}. 
The experiments are conducted using the shadow simulator \cite{shadow2023simulator}. Every node is configured to have 50 Mbps bandwidth, and 100 ms latency is introduced for all edges. The GossipSub related parameters are provided in Table \ref{tab:parameters}.

\begin{figure*}[htb]
    \centering
    \subfigure[Increasing network size]{
        \includegraphics[width=0.315\textwidth]{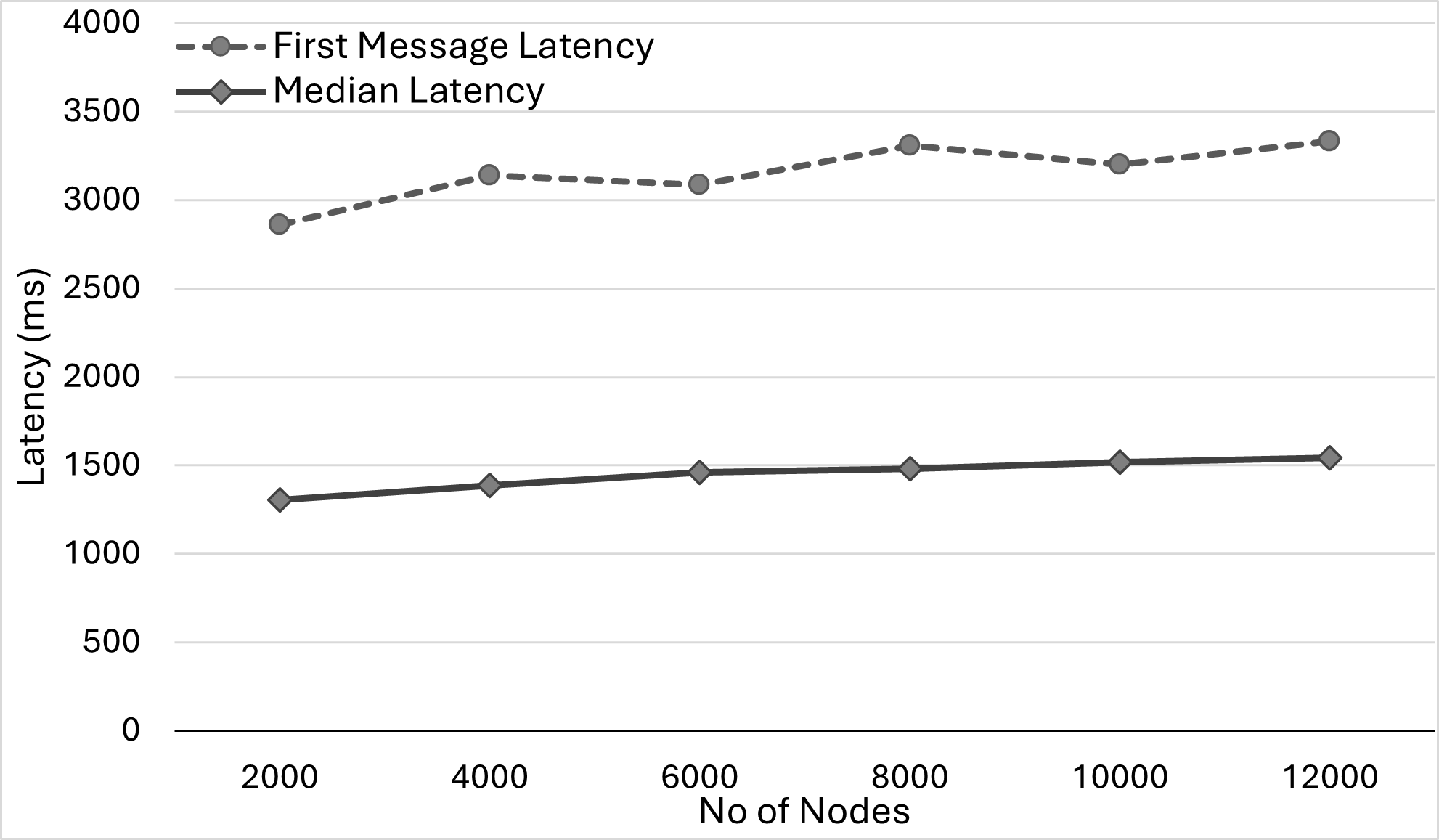}
    }
    \subfigure[Increasing message size]{
        \includegraphics[width=0.315\linewidth]{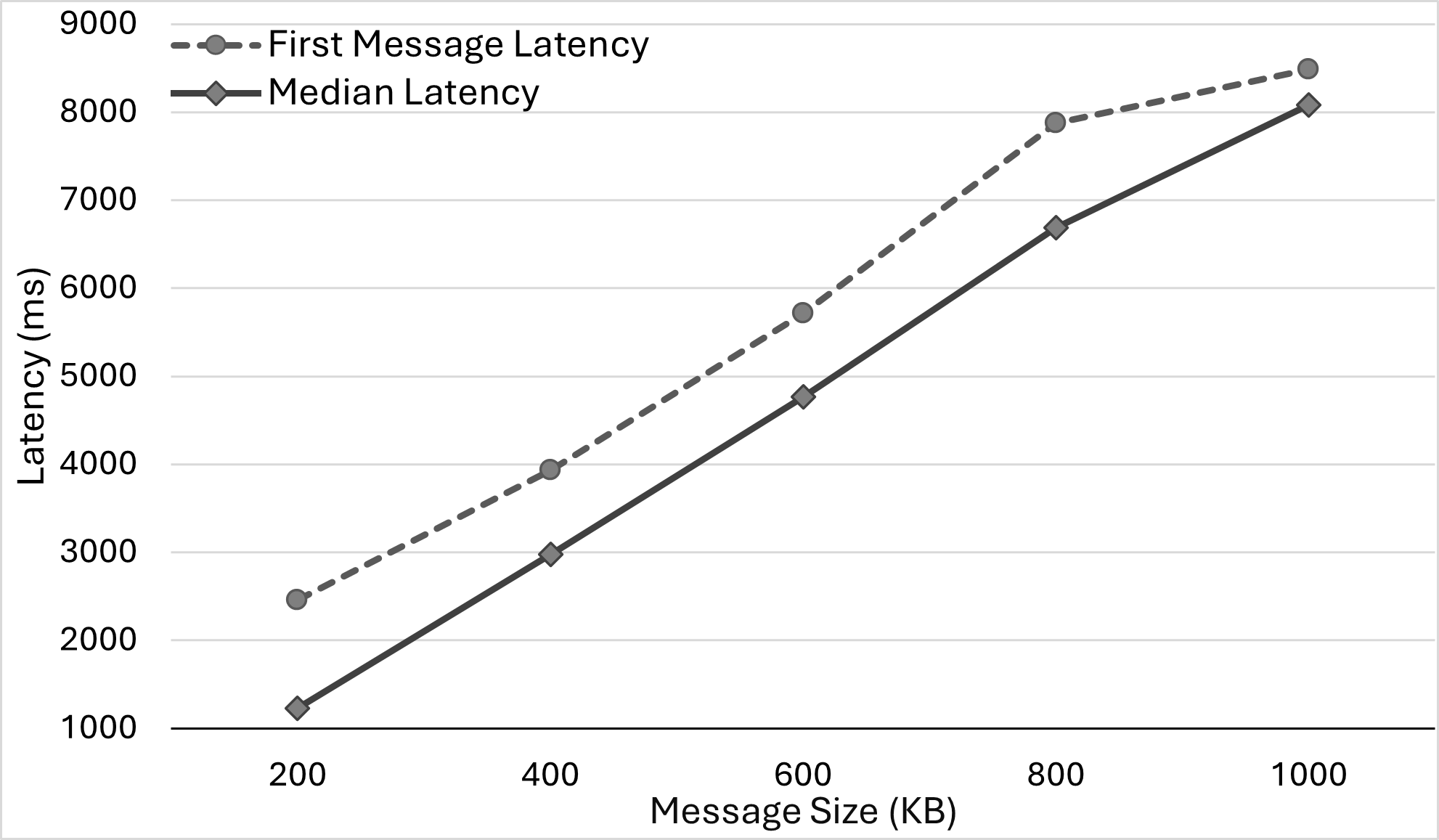}
    }
    \subfigure[Increasing number of publishers]{
        \includegraphics[width=0.315\textwidth]{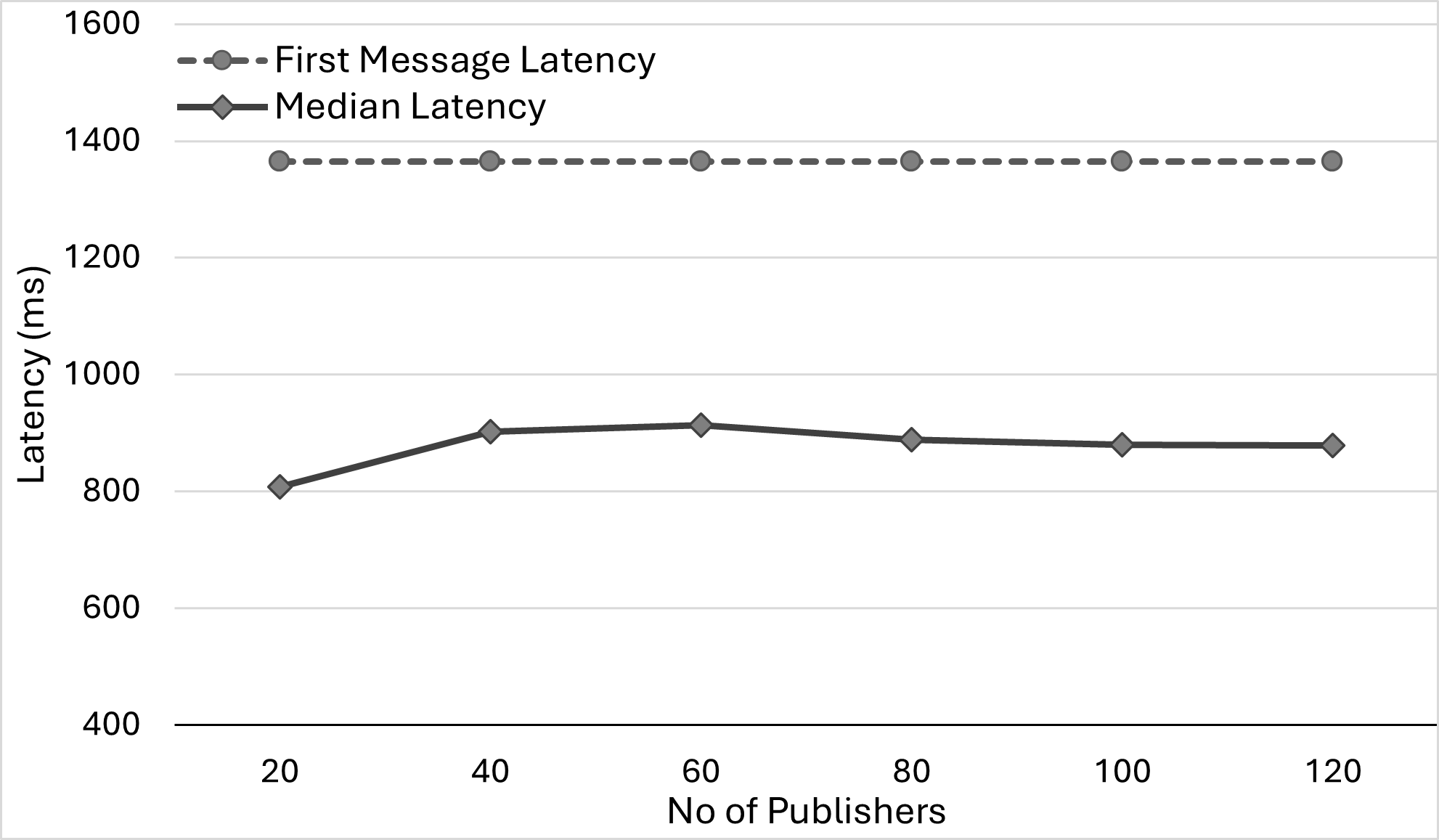}
    }
    \caption{First message latency vs median latency (\emph{TCP $C_{wnd}$})}
    \label{fig:tcp_impact_lat}
\end{figure*}

\begin{figure*}[!t]
    \centering
    \subfigure[Increasing network size (\emph{$L_{200KB}^{2000-12000}$})]{
        \includegraphics[width=0.315\textwidth]{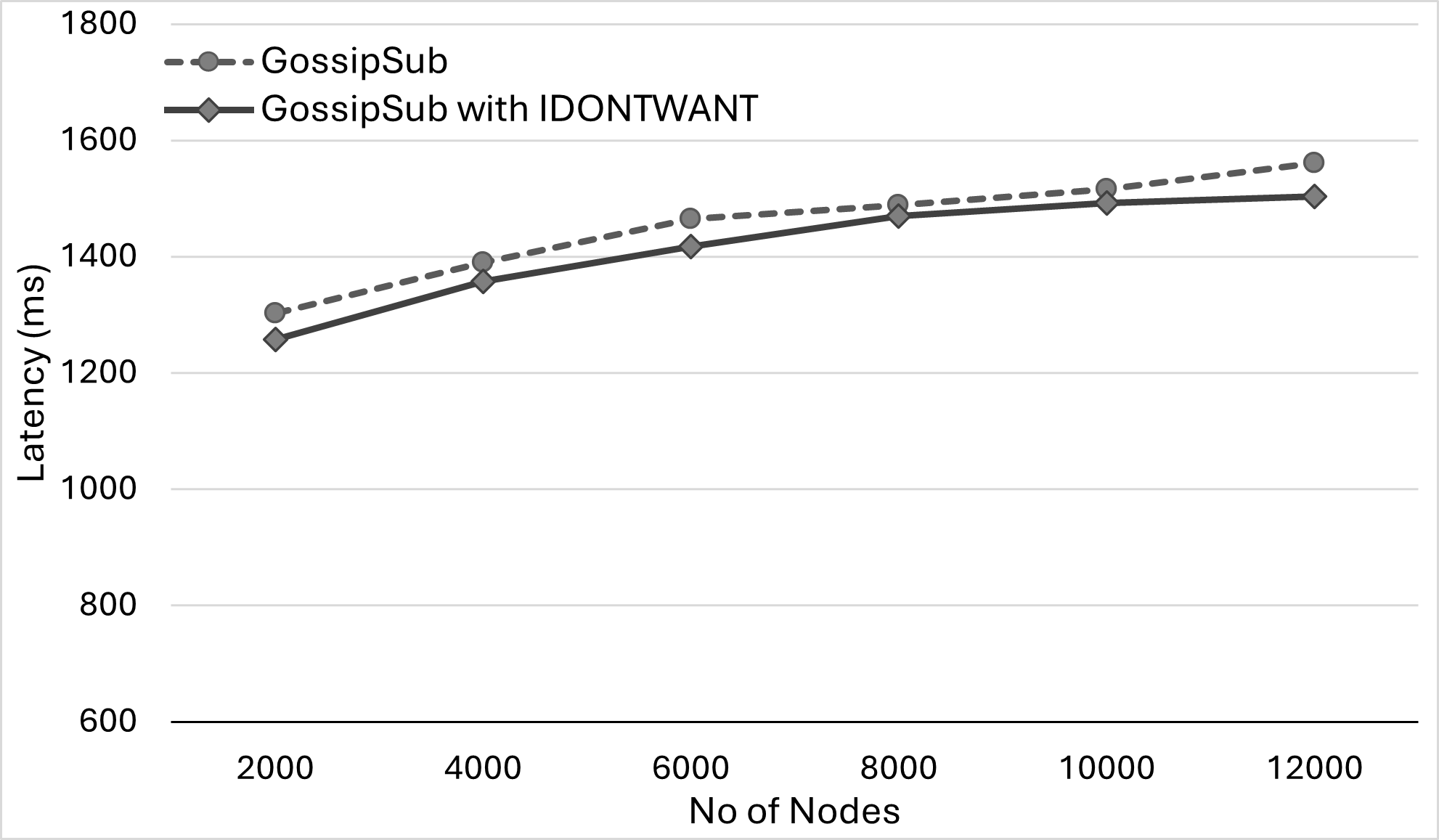}
    }
    \subfigure[Increasing message size (\emph{$L_{200-1000KB}^{1000}$})]{
        \includegraphics[width=0.315\linewidth]{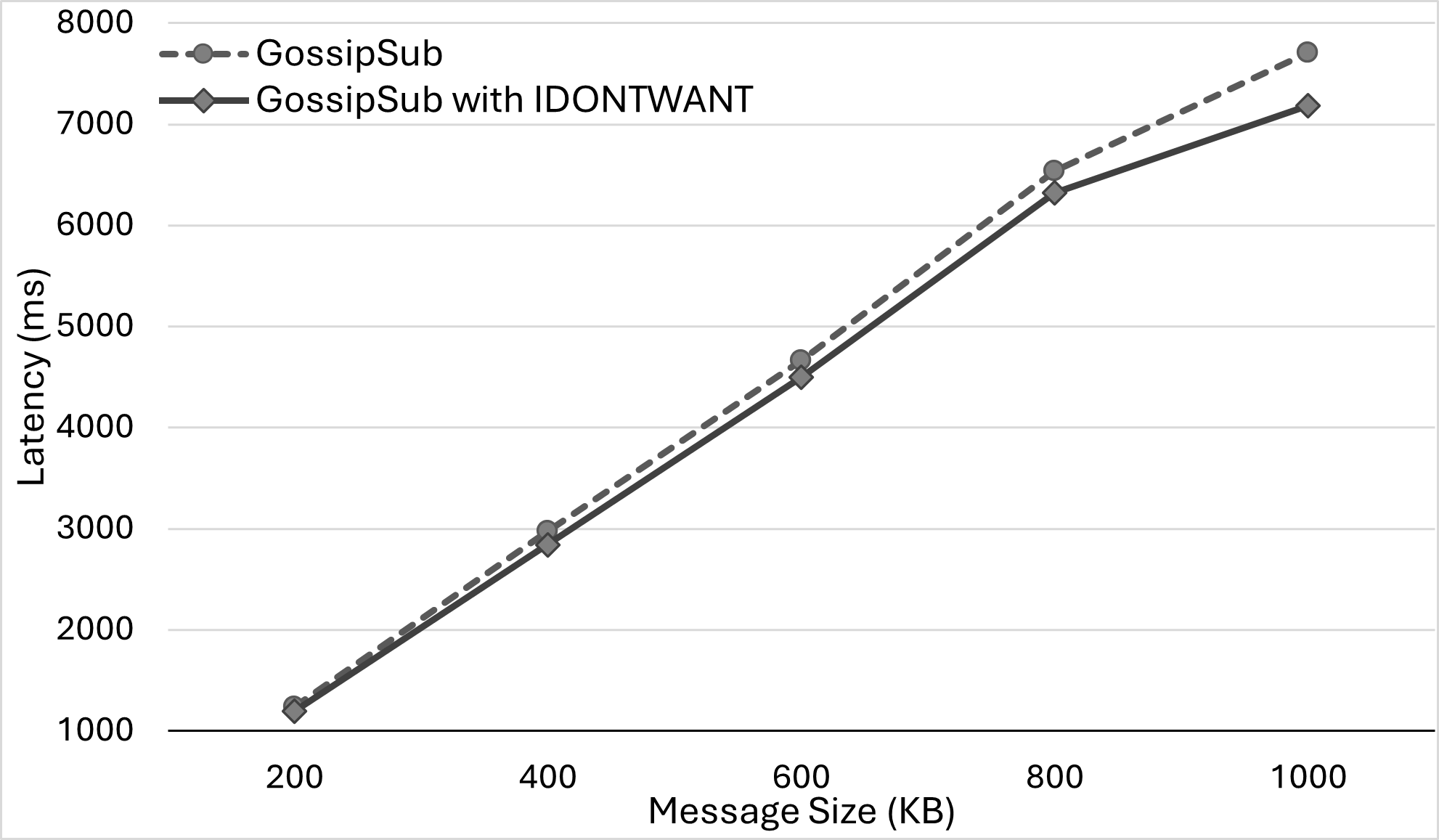}
    }
    \subfigure[Increasing number of publishers (\emph{$L_{50KB}^{1000}$})]{
        \includegraphics[width=0.315\linewidth]{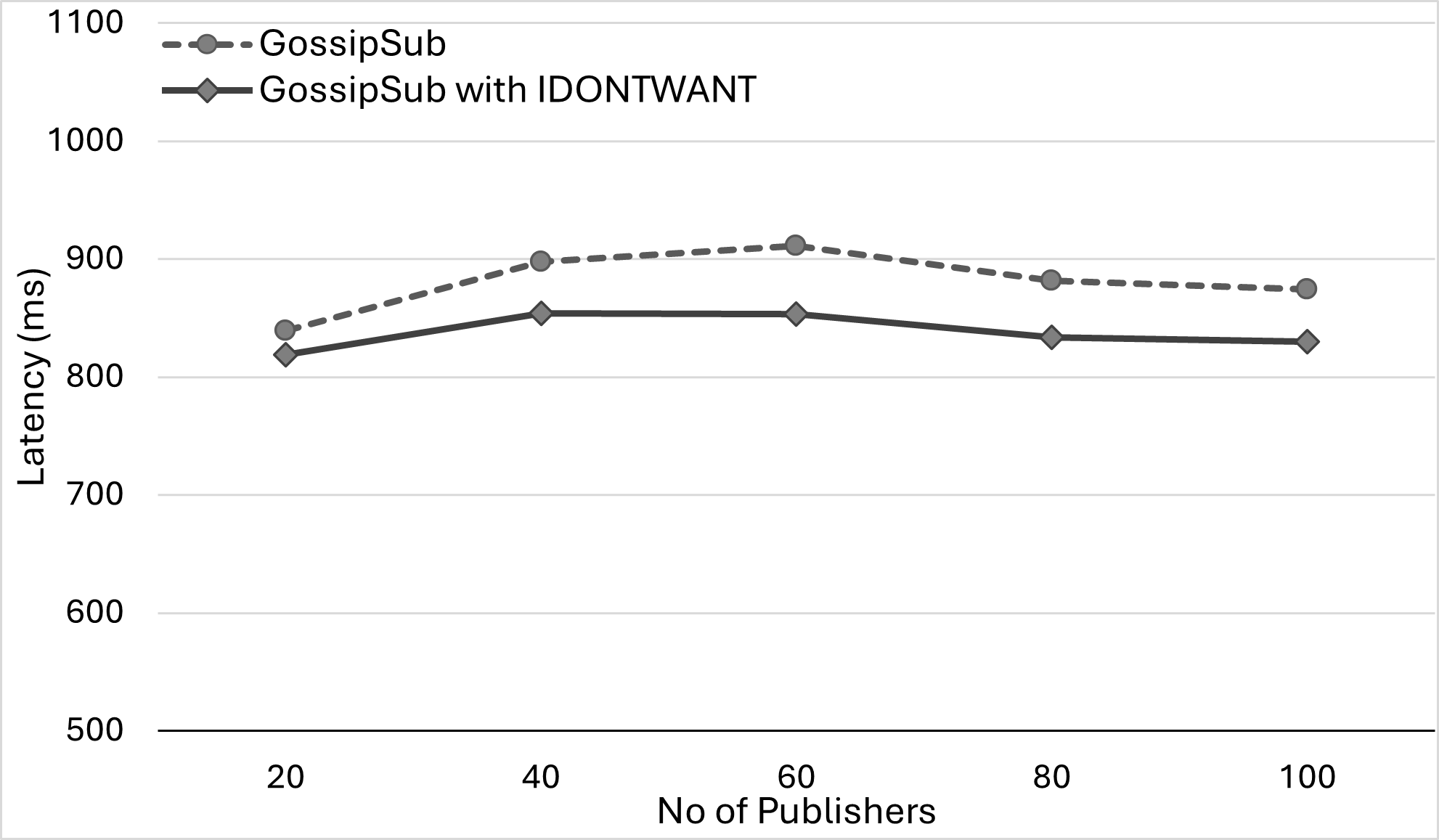}
    }
    \subfigure[Increasing network size (\emph{$B_N$})]{
        \includegraphics[width=0.315\linewidth]{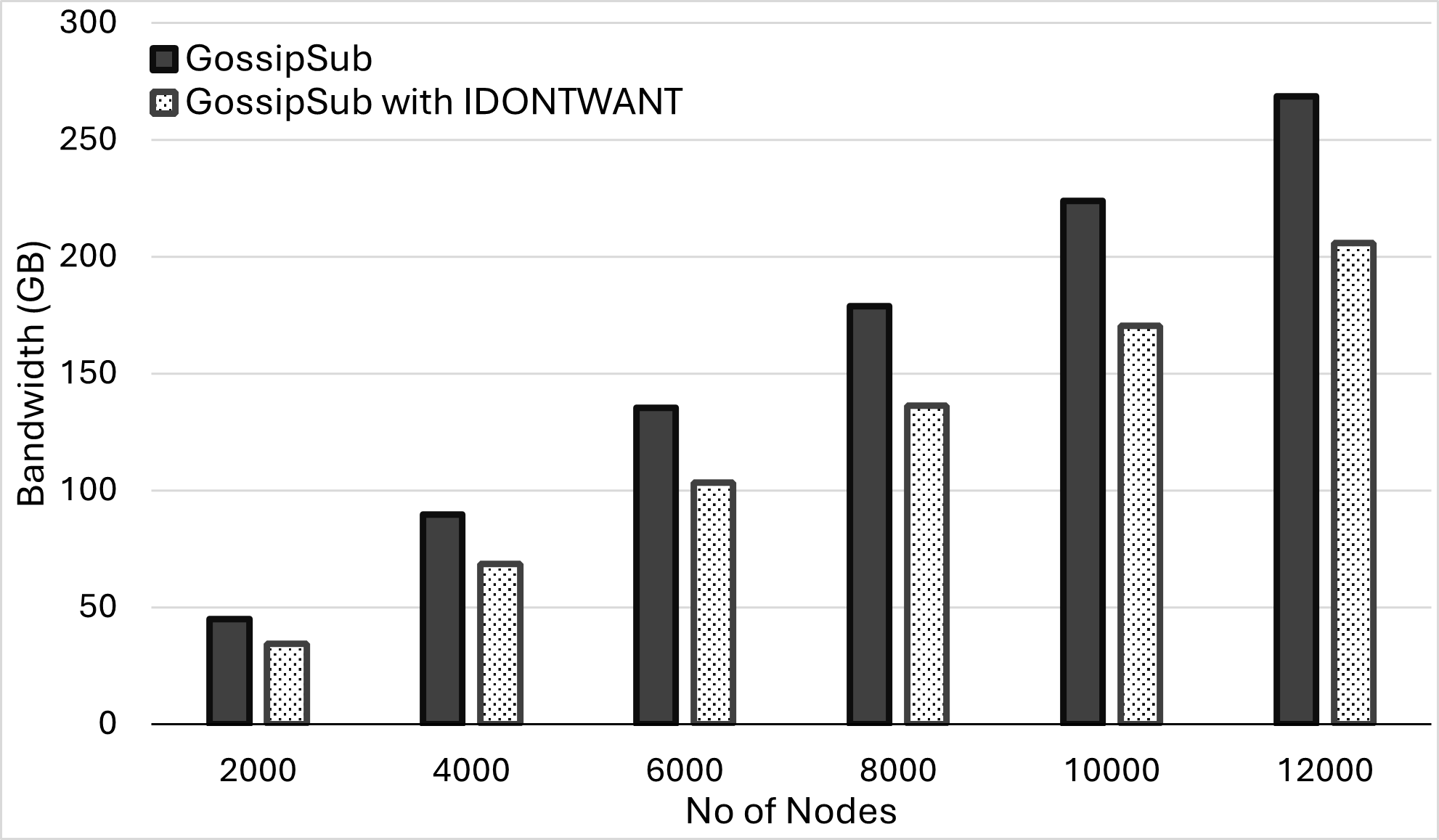}
    }
    \subfigure[Increasing message size (\emph{$B_N$})]{
        \includegraphics[width=0.315\linewidth]{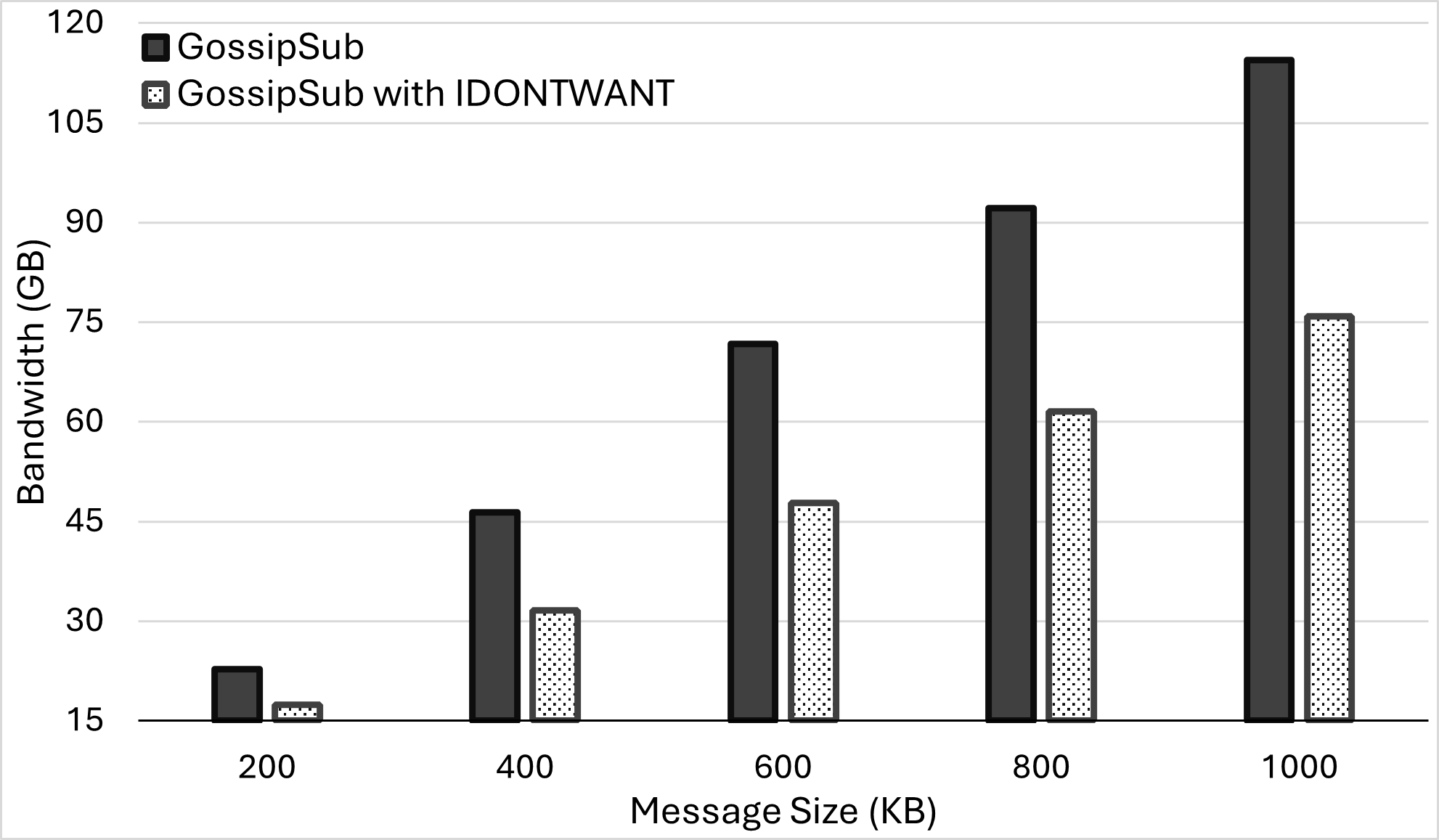}
    }
    \subfigure[Increasing number of publishers (\emph{$B_N$})]{
        \includegraphics[width=0.315\linewidth]{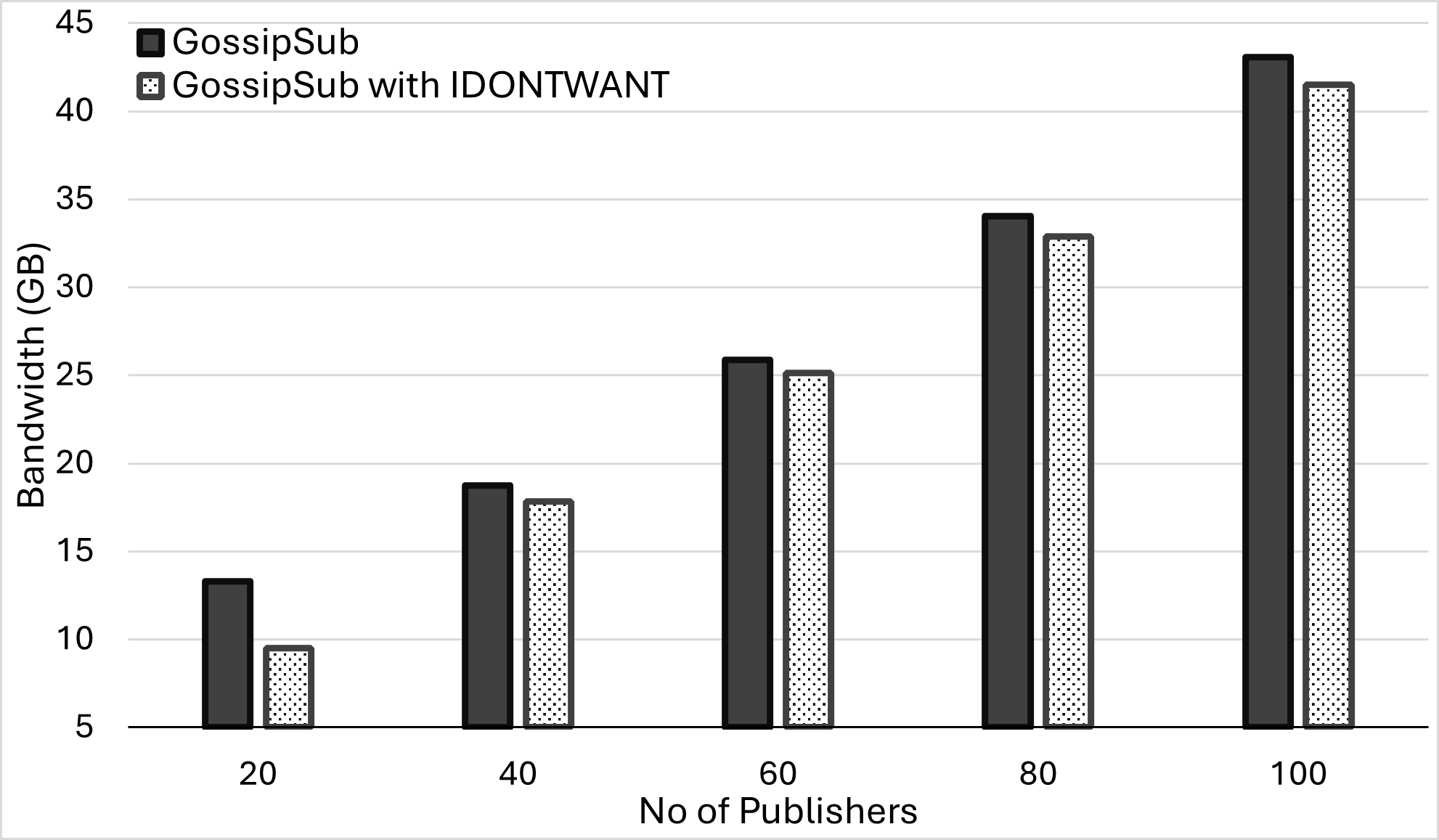}
    }
    
    \caption{IDONTWANT message impact: average latency (\emph{$L_{cov}^{100}$}) and bandwidth utilization (\emph{$B_N$})}
    \label{fig:idontwant_message}
\end{figure*}

We first discuss the network-wide message dissemination time $L_{cov}^{100}$. The theoretical time needed to transfer a block to the entire network can be estimated as $L_{cov}^{100} \approx (\tau_p + \tau_{tx}) \times H$, where $\tau_p$ and $\tau_{tx}$ represent single-hop propagation and transmit times respectively. $H$ is the length of the longest path calculated as $H = \lceil \frac{\log N}{\log D} \rceil$ with $N$ being the number of nodes and $D$ being the average node degree in the full-message mesh. Assuming $D = 8$, an average link latency of 100 ms, and a peer bandwidth of 50 Mbps, we can estimate the time required for disseminating a 1MB message in a network with 1000 peers as $L_{1MB}^{1000} \approx (100 + \frac{1MB \times 8}{50 Mbps}) \times 4 = 5520 ms$. However, this estimate depends on the randomness and D-regularity of the full-message mesh. Duplicate receptions in the earlier phase of message propagation (or smaller D) can lead to an increased number of rounds (hops), resulting in higher latency. The waiting time in outgoing message queues at optimal path peers also contributes to latency.

TCP congestion avoidance mechanisms also have a noticeable impact on message dissemination latency. The congestion avoidance algorithms usually limit maximum in-flight bytes in a round trip time (RTT) based on the $C_{wnd}$. A lower $C_{wnd}$ in a newly established (cold) TCP connection may result in a much longer message transmission time. However, $C_{wnd}$ rises with the data transfer. Consequently, sending the same message through a cold connection takes longer. The message transfer time lowers as the $C_{wnd}$ grows. That is why a much higher message dissemination latency is observed for first message transfers in GossipSub, which decreases for the subsequent messages as depicted in Fig. \ref{fig:tcp_impact_lat}. This also indicates that sending a large message through less frequent connections, such as IWANT replies or floodpublish, may take longer than the usual transmission time. For the rest of the article, we consider the first two messages as warm-up messages and exclude them from all computations except for bandwidth utilization.

\begin{figure*}[!t]
    \centering
    \subfigure[Increasing network size (\emph{$L_{200KB}^{2000-12000}$})]{
        \includegraphics[width=0.315\textwidth]{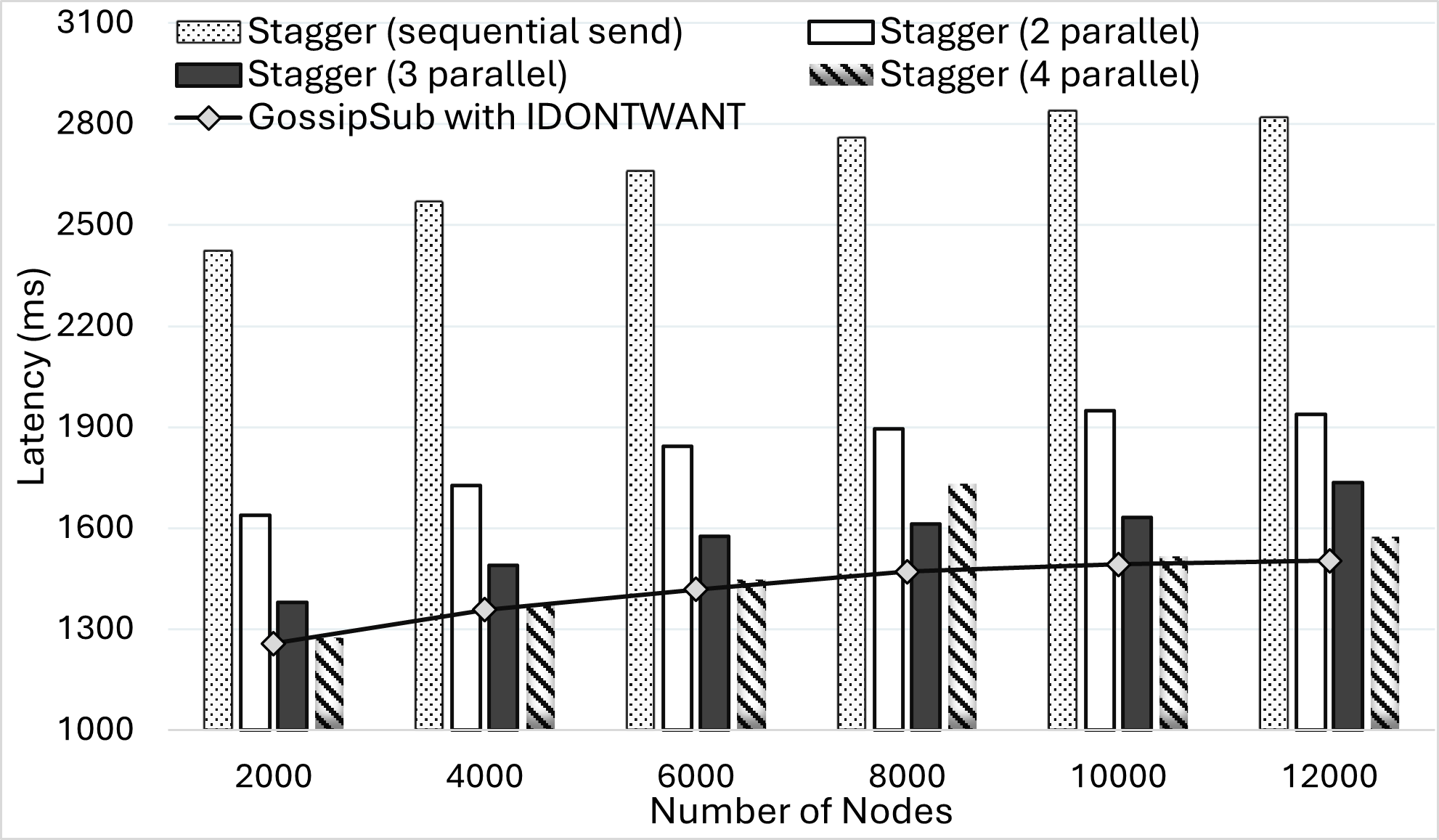}
    }
    \subfigure[Increasing message size (\emph{$L_{200-1000KB}^{1000}$})]{
        \includegraphics[width=0.315\linewidth]{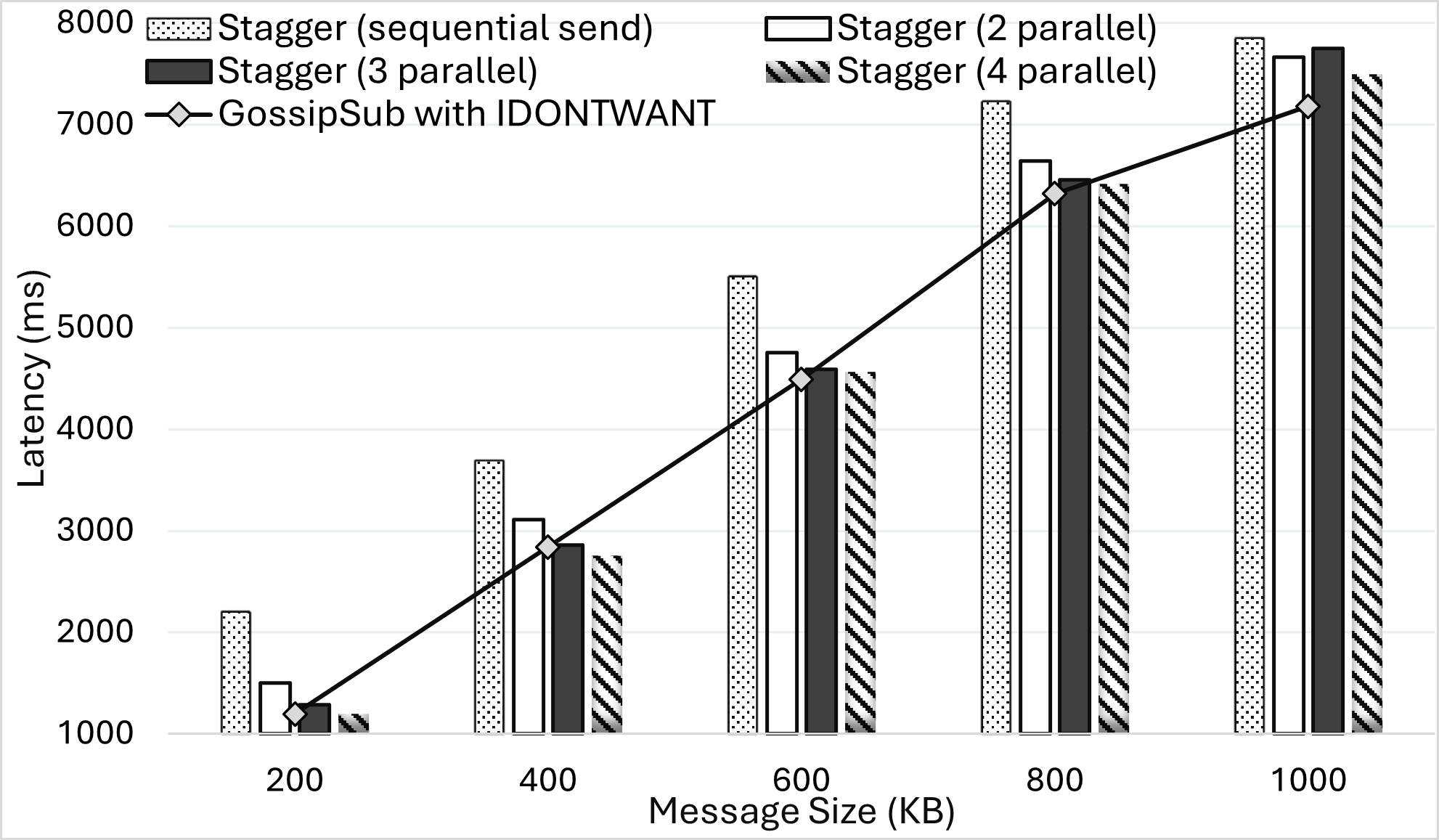}
    }
    \subfigure[Increasing number of publishers (\emph{$L_{50KB}^{1000}$})]{
        \includegraphics[width=0.315\linewidth]{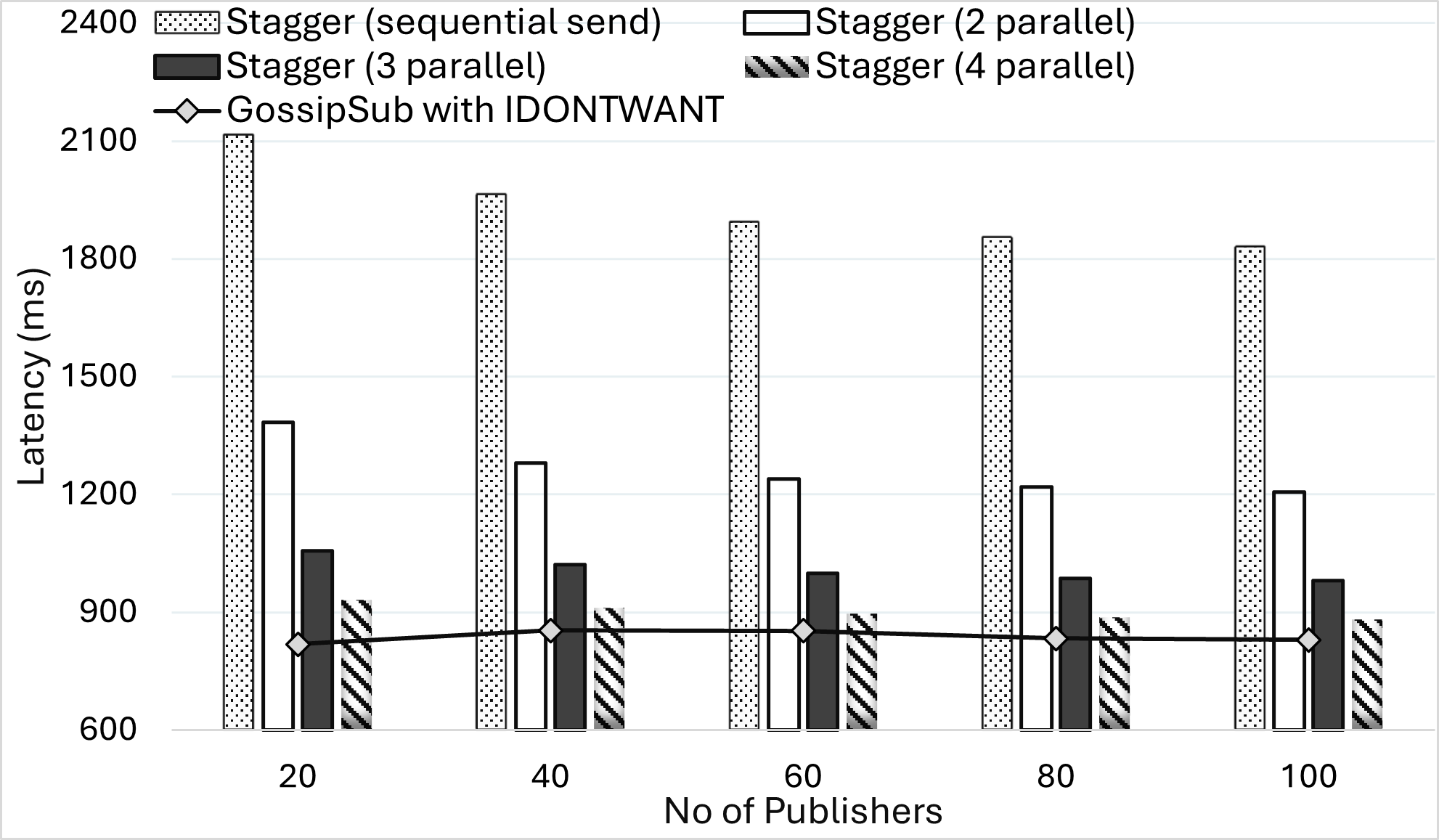}
    }

    \subfigure[Increasing network size (\emph{$B_N$})]{
        \includegraphics[width=0.315\linewidth]{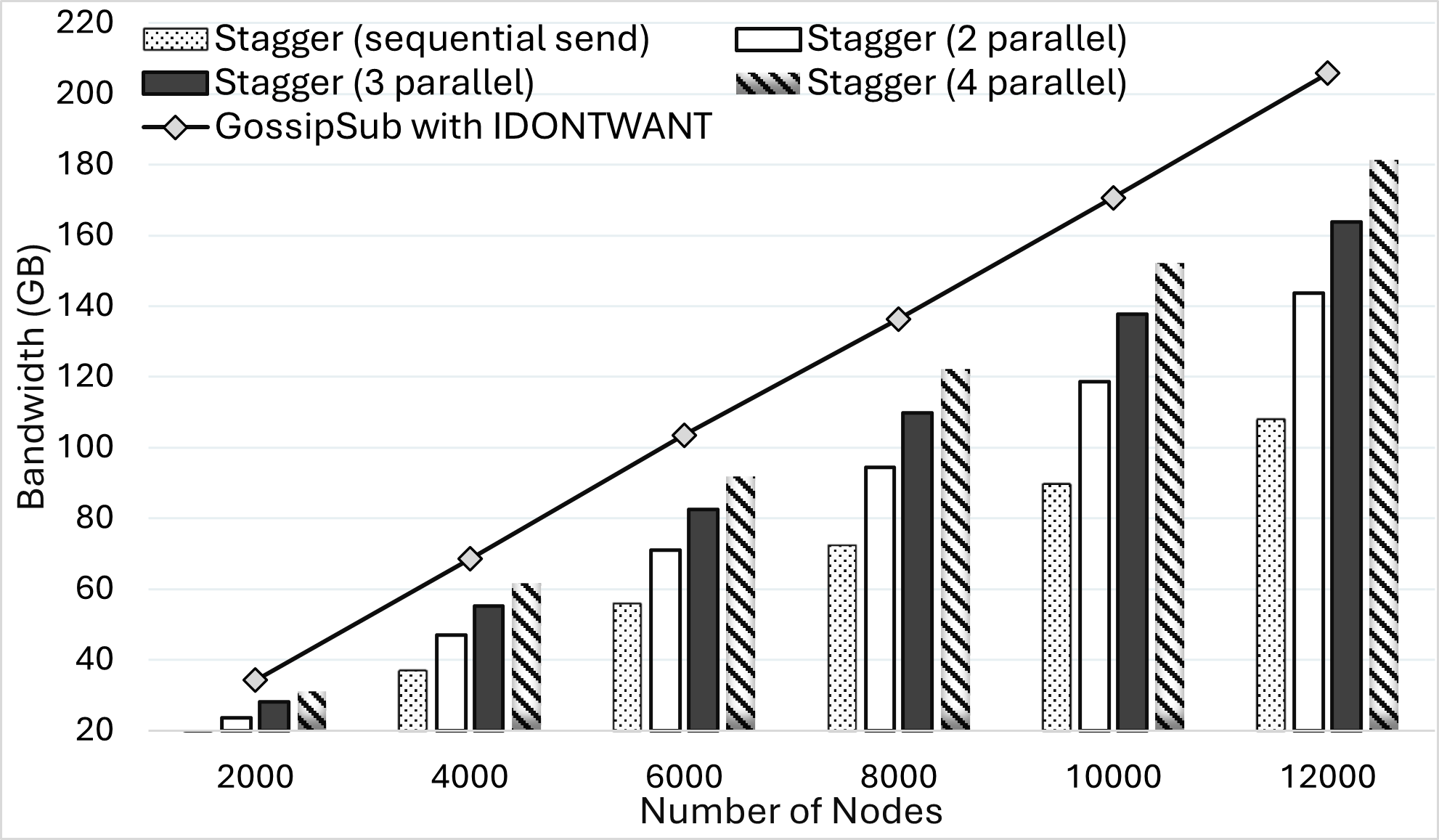}
    }
    \subfigure[Increasing message size (\emph{$B_N$})]{
        \includegraphics[width=0.315\linewidth]{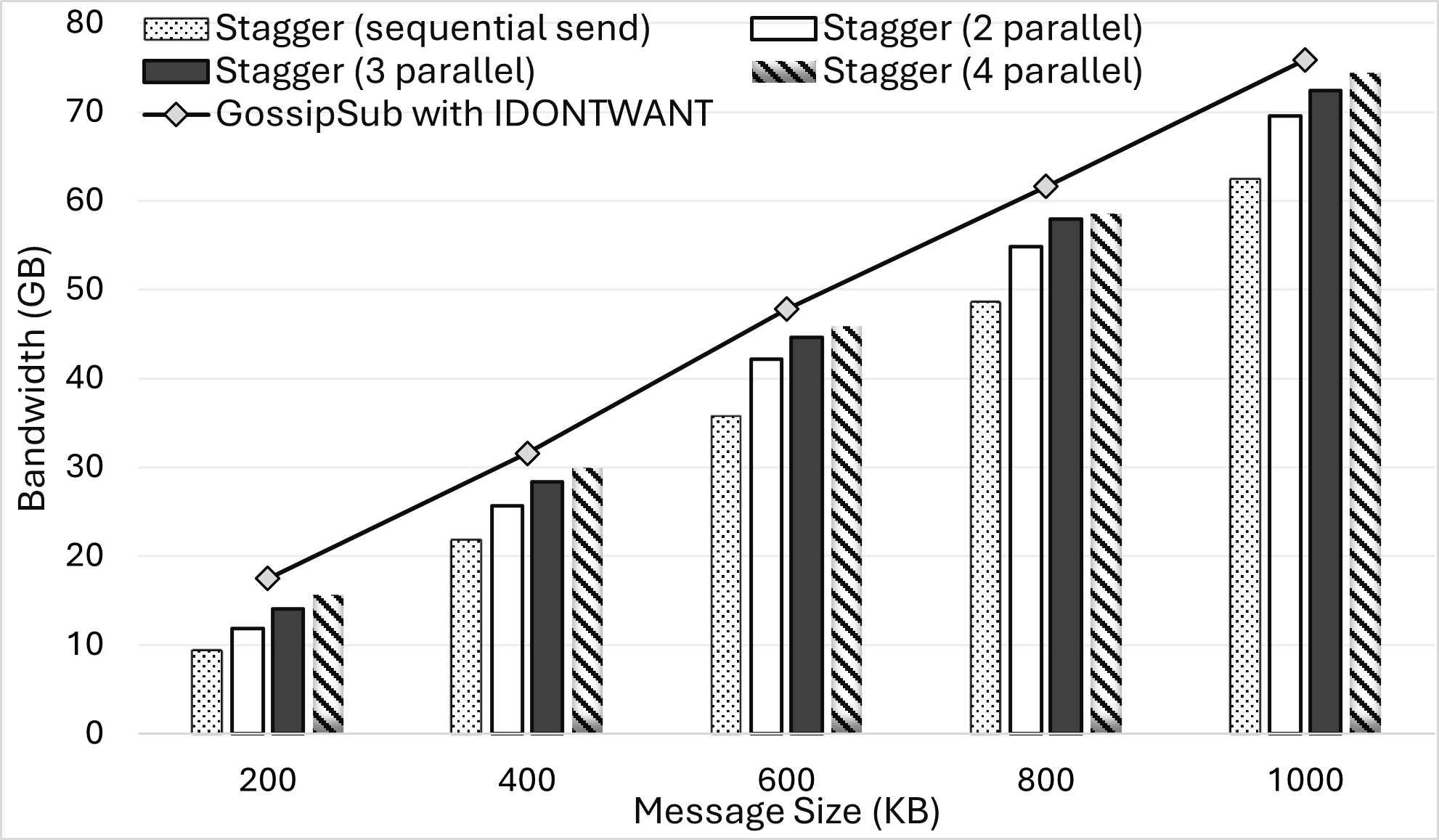}
    }
    \subfigure[Increasing number of publishers (\emph{$B_N$})]{
        \includegraphics[width=0.315\linewidth]{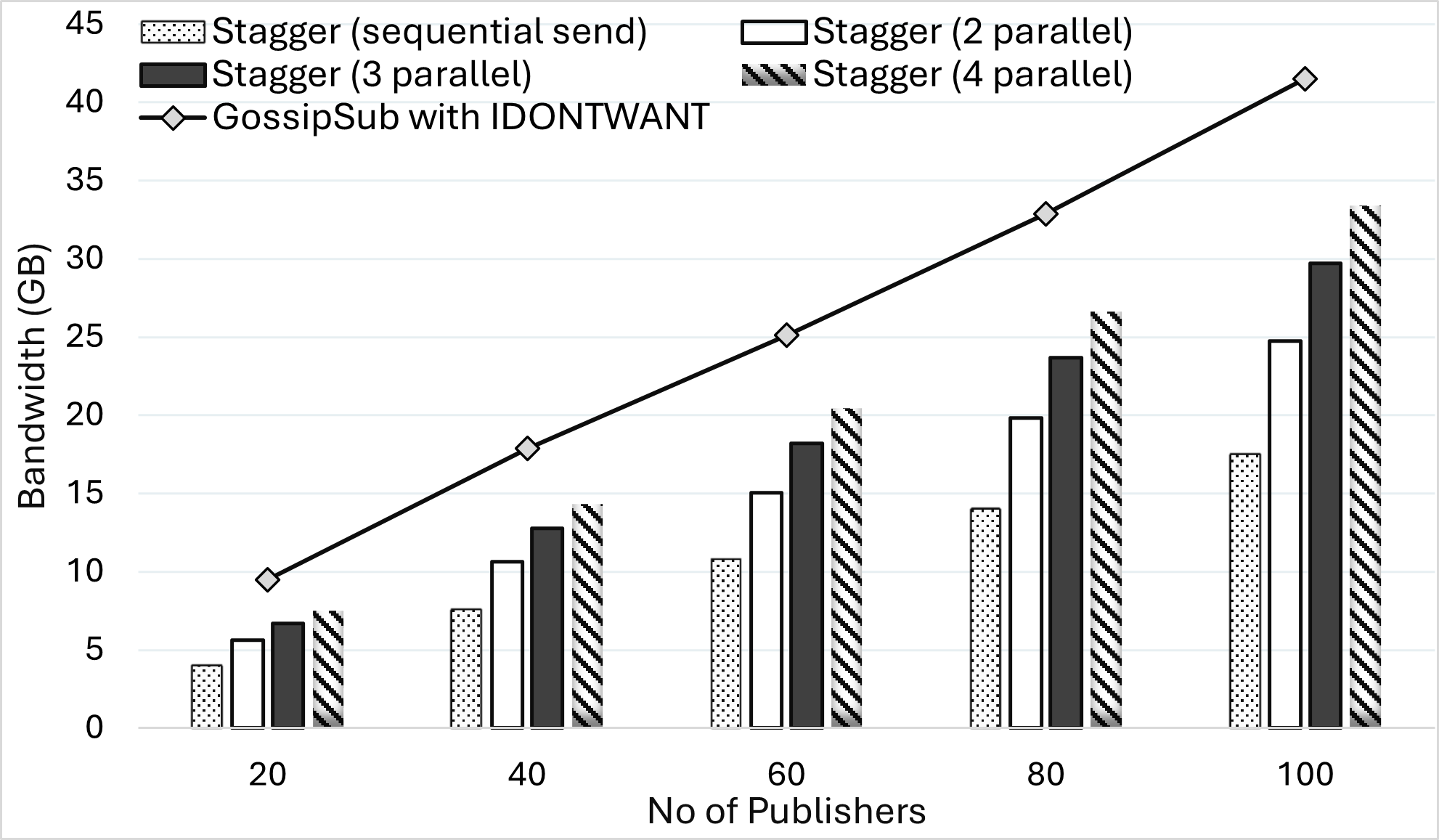}
    }

    \subfigure[Increasing network size (\emph{IWANT requests})]{
        \includegraphics[width=0.315\linewidth]{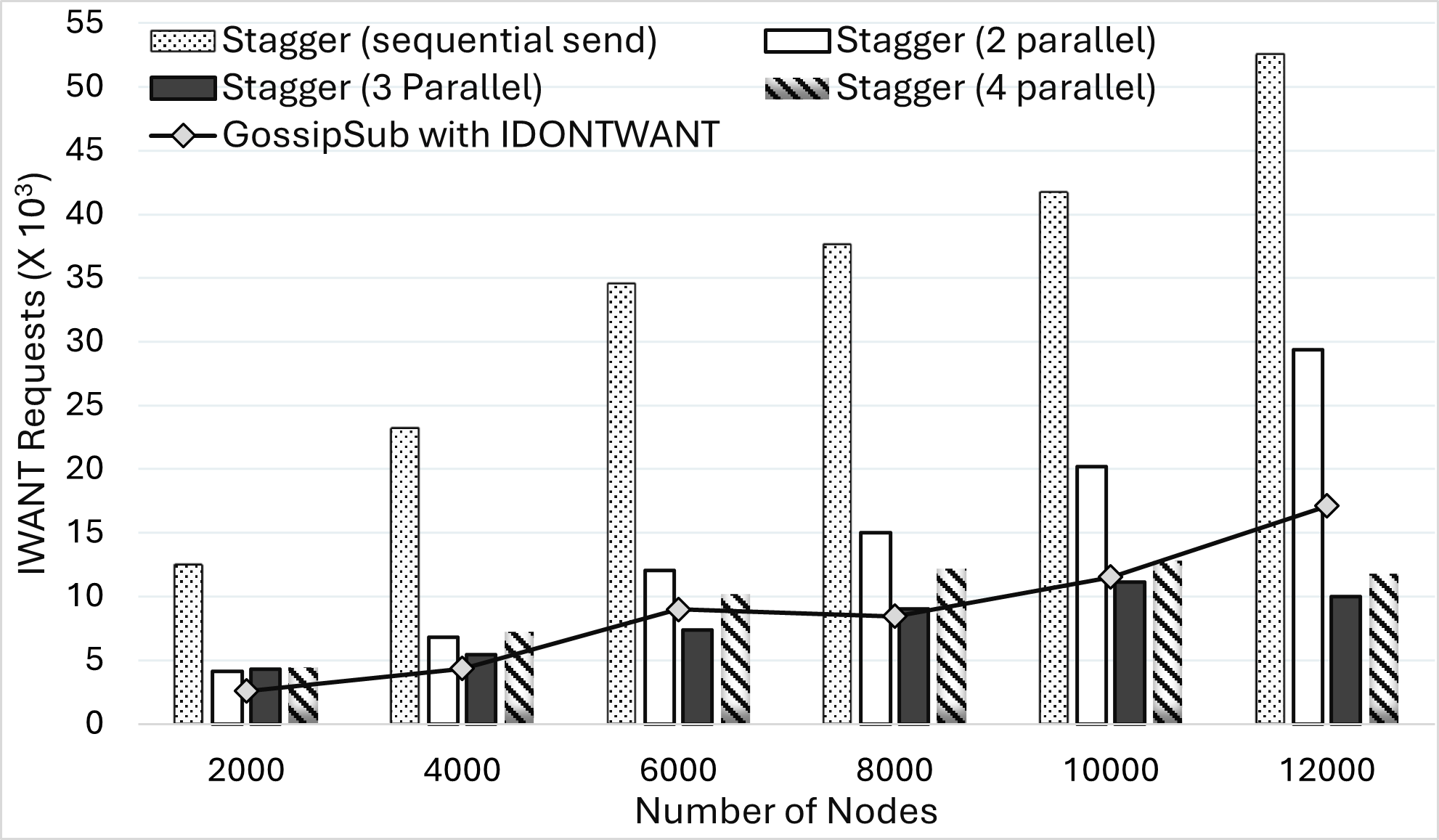}
    }
    \subfigure[Increasing message size (\emph{IWANT requests})]{
        \includegraphics[width=0.315\linewidth]{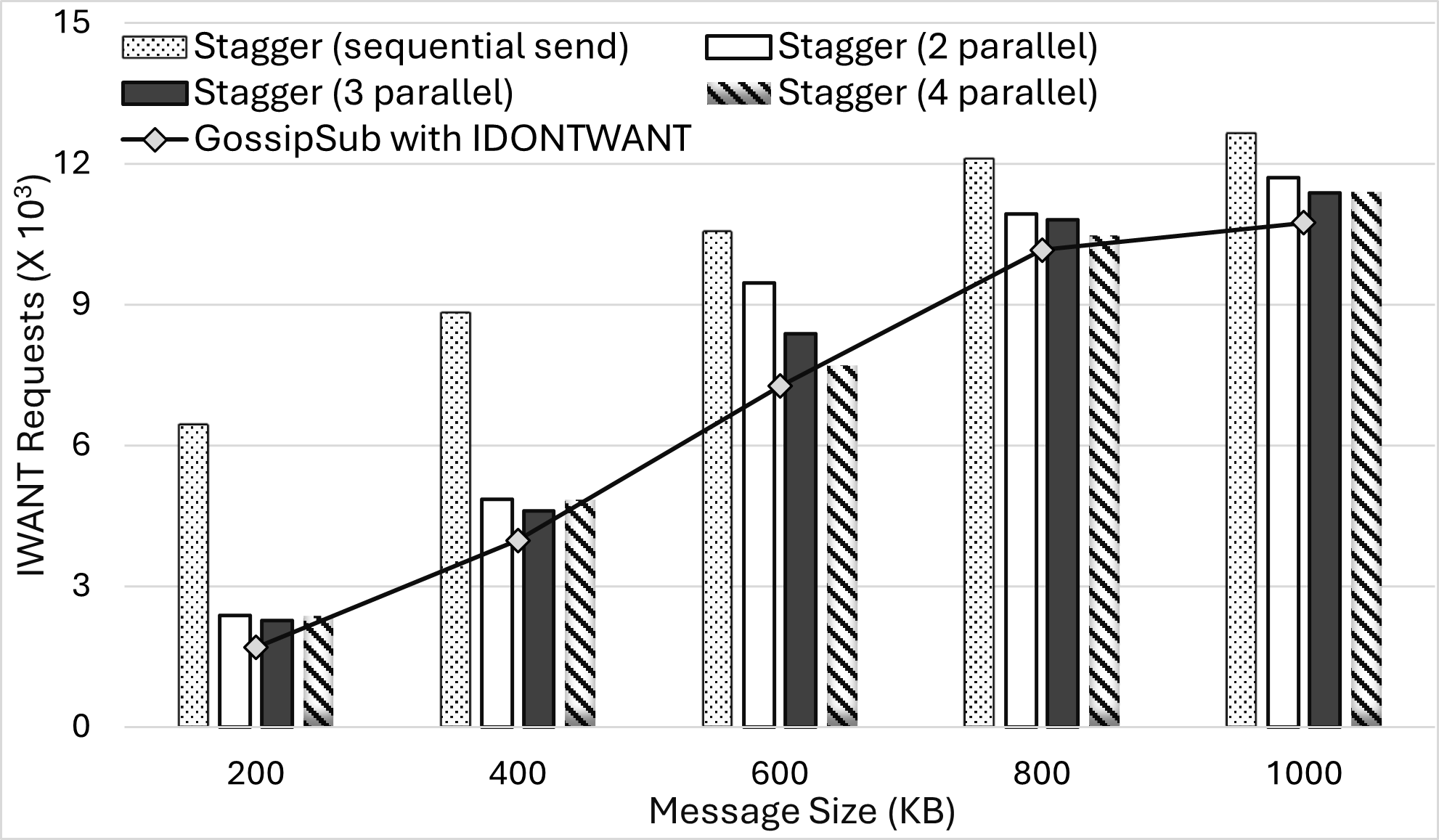}
    }
    \subfigure[Increasing publishers (\emph{IWANT requests})]{
        \includegraphics[width=0.315\linewidth]{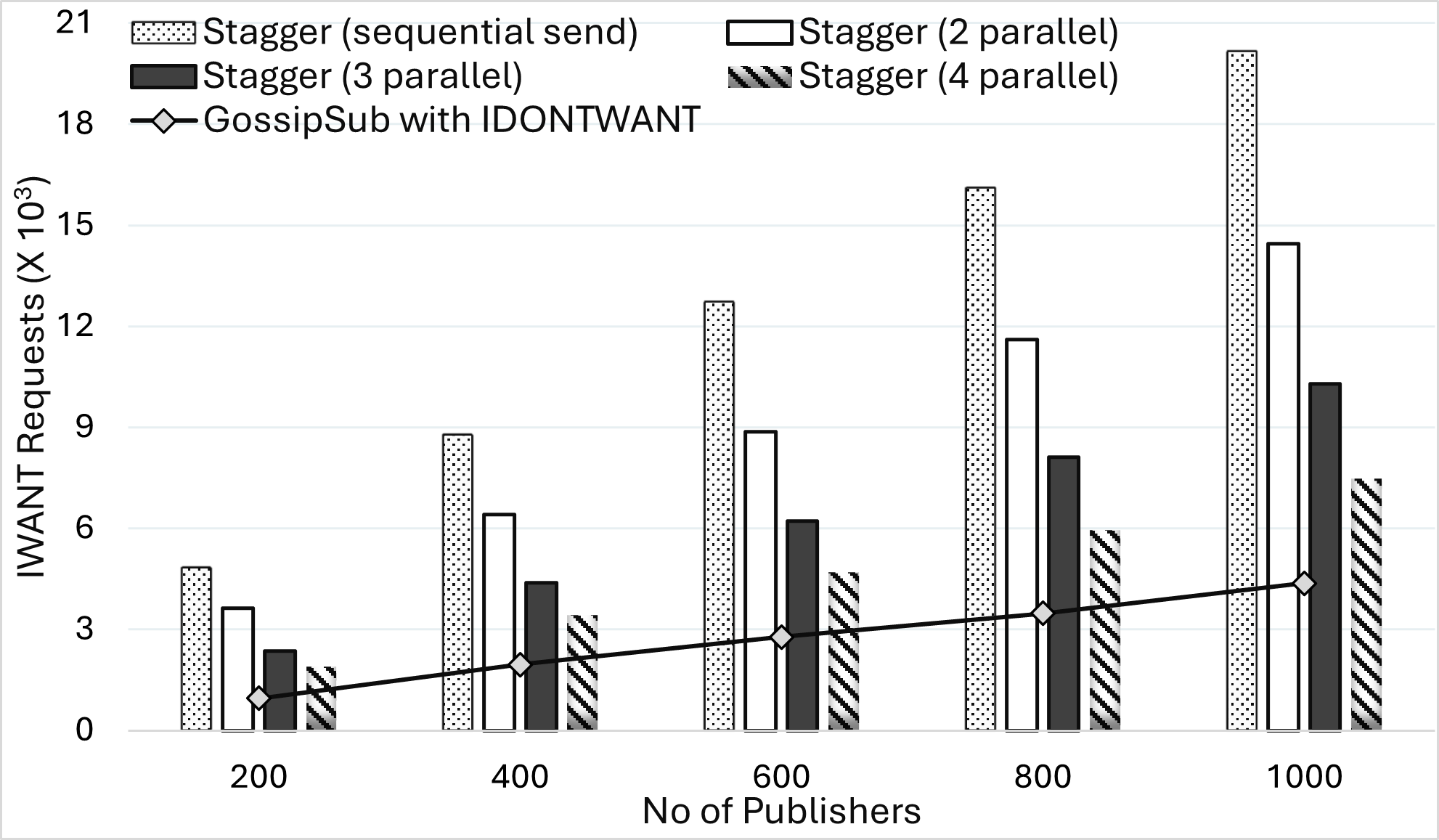}
    }
    
    \caption{Message-staggering with IDONTWANT messages: average Latency (\emph{$L_{cov}^{100}$}),  bandwidth utilization (\emph{$B_N$}), and the number of IWANT requests}
    \label{fig:stagger_idontwant}

    \vspace{0.1cm}
     
    \scriptsize \raggedright IDONTWANT messages are enabled for all the protocols depicted in this figure. We use GossipSub with IDONTWANT message support (GossipSub v1.2) as a baseline protocol. For message-staggering, transmissions are carried out sequentially, followed by parallel transmissions to groups of two, three, and four peers. We use a 200 ms timeout interval (stagger interval) for message-staggering, i.e., If a transmission does not complete within 200 ms, we start relaying to the next group of peers.
\end{figure*}

The impact of IDONTWANT messages on latency and bandwidth reduction is illustrated in Fig. \ref{fig:idontwant_message}. IDONTWANT messages lead to significant bandwidth savings, proportional to the message sizes. However, to maximize bandwidth reduction, a peer must receive and process IDONTWANT notifications for a message before it starts relaying that message. It is important to note that receiving a large message can take considerable time. During that time, peers are not able to inform their mesh members about the ongoing message reception. Addressing this issue can further enhance bandwidth savings associated with IDONTWANT messages. On the other hand, an almost similar latency is observed with IDONTWANT messages in Fig. \ref{fig:idontwant_message}(a)-(c). Only a small reduction in $L_{cov}^{100}$ can be attributed to IDONTWANT messages at high traffic volumes. This is due to the fact that IDONTWANT messages can reduce $L_{cov}^{100}$ only when they alleviate the outgoing message queue sizes at the optimal path peers. However, these peers are early receivers and typically do not receive IDONTWANT notifications for most messages. A significant rise in $L_{200-1000KB}^{1000}$ (Fig. \ref{fig:idontwant_message}-(b)) is attributed to the fact that increasing message size also proportionally increases message transmission time. This additional store-and-forward delay accumulates across the message propagation path. 

Message-staggering can be an effective approach to reduce store-and-forward delay. It can minimize message download time for individual peers by concentrating the sender's bandwidth for a single transmission. As a result, more peers start spreading the message during early stages of message propagation, while the original sender continues to relay the message to the remaining mesh members. However, message-staggering indicates a noticeable increase in latency in Fig. \ref{fig:stagger_idontwant}(a)-(c). This rise in latency can be attributed to multiple factors: 1) A sender may block for acknowledgments during large message transfers, as TCP limits the maximum in-flight bytes based on $C_{wnd}$. As a result, a smaller $C_{wnd}$ or a higher link latency may diminish the benefits of sequential sending. Table \ref{tab:warmup_latency} depicts that message-staggering over smaller link latencies with warm TCP connections yields much lower $L_{200-1000KB}^{1000}$ for large message transfers. 2) Early message receivers get many IWANT requests, as they are among the few peers announcing a new message ID in their IHAVE announcements. The steep rise in the number of IWANT requests in Fig. \ref{fig:stagger_idontwant}(g)-(i) supports this assumption. Notably, many of these IWANT requests may be duplicates (GossipSub permits sending multiple IWANT requests for the same message ID) or may have been issued by peers already receiving the same message from other senders. 3) A slow receiver may block a fast sender. Additionally, if multiple senders select the same receiver during the early stages of message propagation, the message spread in the network can also slow down. A straightforward remedy is to choose a small subset of mesh peers to relay a message rather than sending it to one receiver at a time. Fig. \ref{fig:stagger_idontwant}(a)-(c) highlights that message-staggering with 3-4 parallel sends noticeably reduces latency. 4) Sequential message forwarding overhead can also diminish the benefits of message-staggering for small messages. However, achieving optimal message-staggering is challenging, and different implementation strategies may distinctly impact $L_{cov}^{100}$. 

\begin{figure*}[!t]
    \centering
    \subfigure[Increasing network size (\emph{$L_{200KB}^{2000-12000}$})]{
        \includegraphics[width=0.315\textwidth]{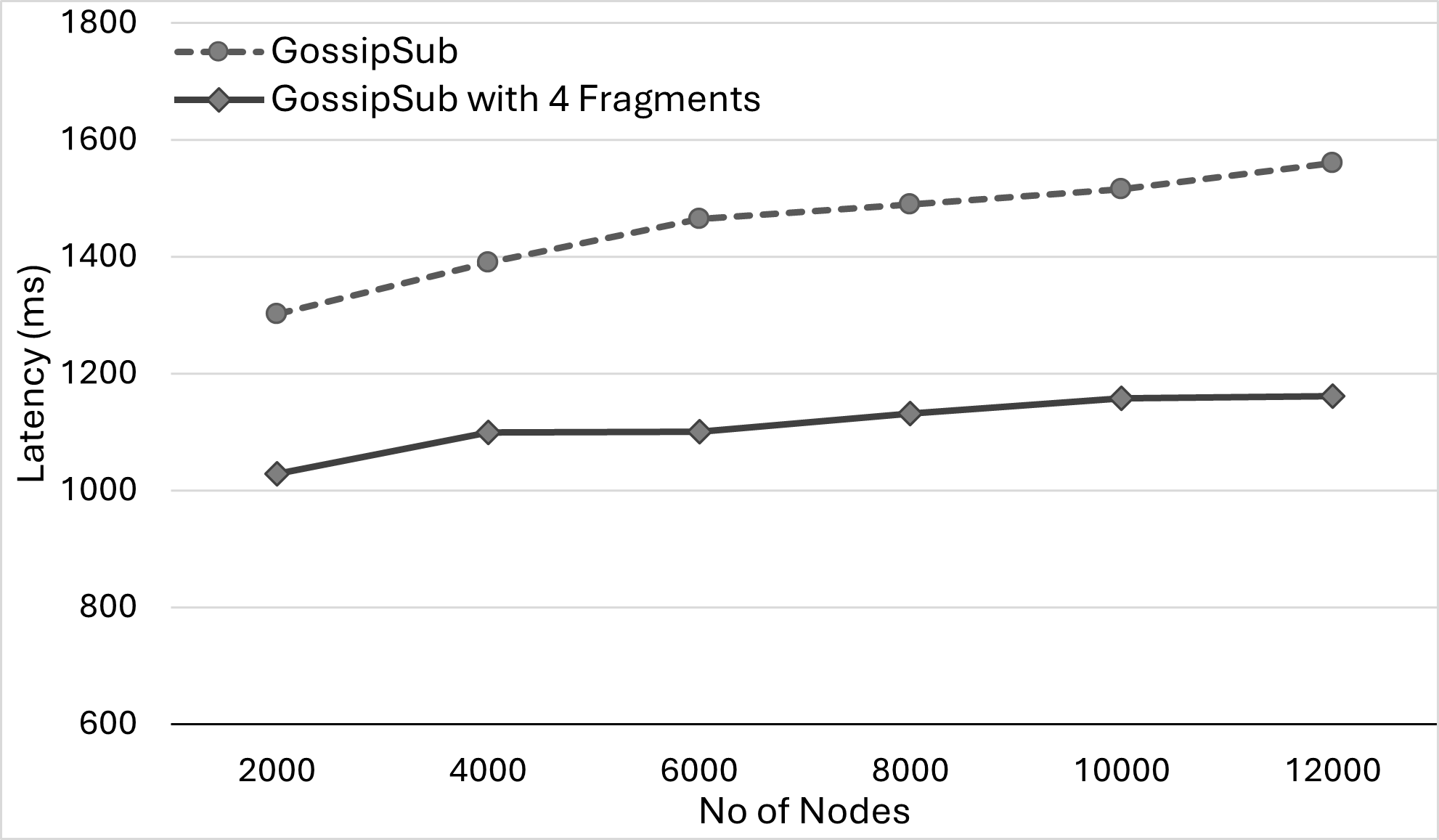}
    }
    \subfigure[Increasing message size (\emph{$L_{200-1000KB}^{1000}$})]{
        \includegraphics[width=0.315\linewidth]{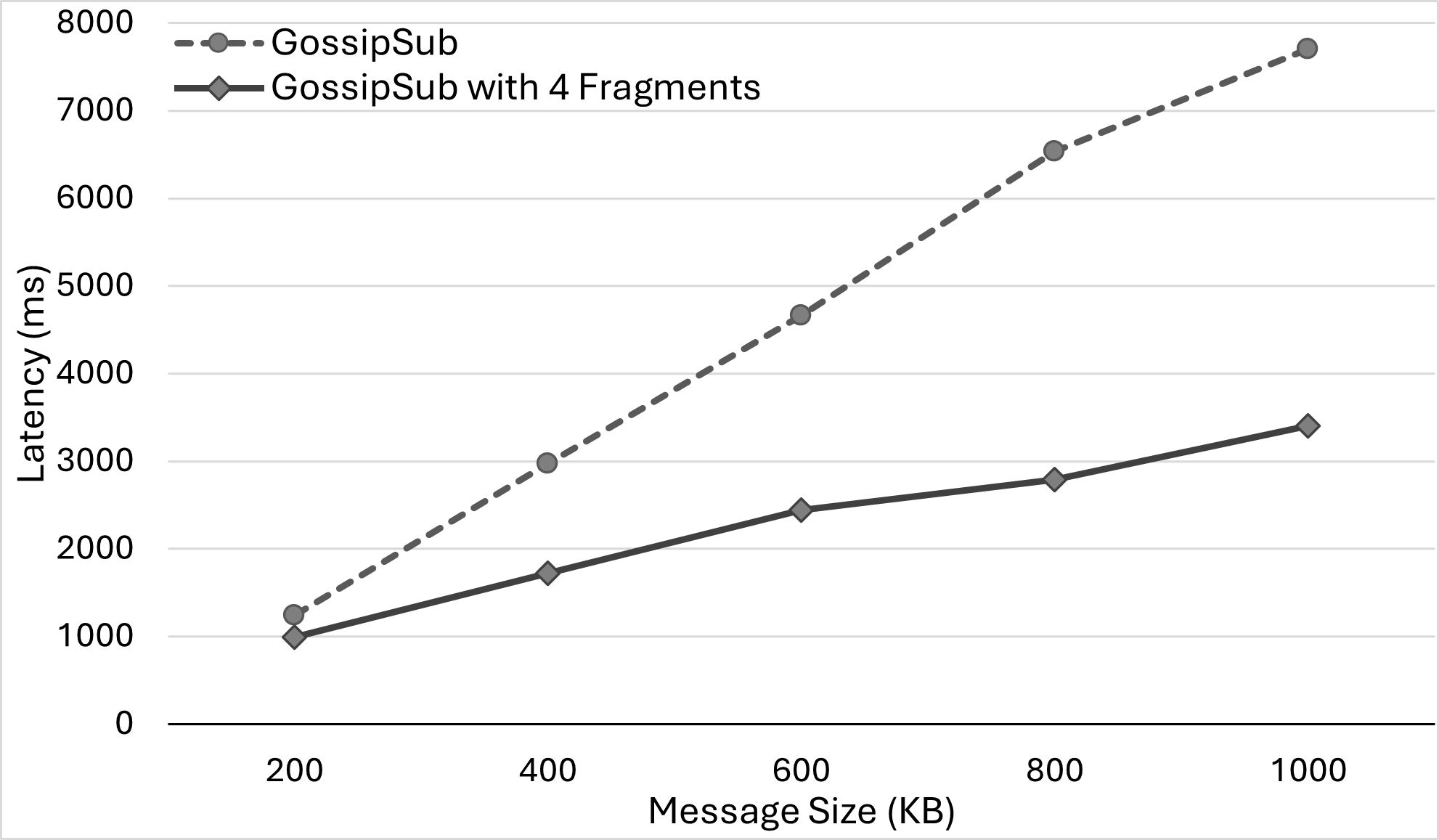}
    }
    \subfigure[Increasing number of publishers (\emph{$L_{50KB}^{1000}$})]{
        \includegraphics[width=0.315\linewidth]{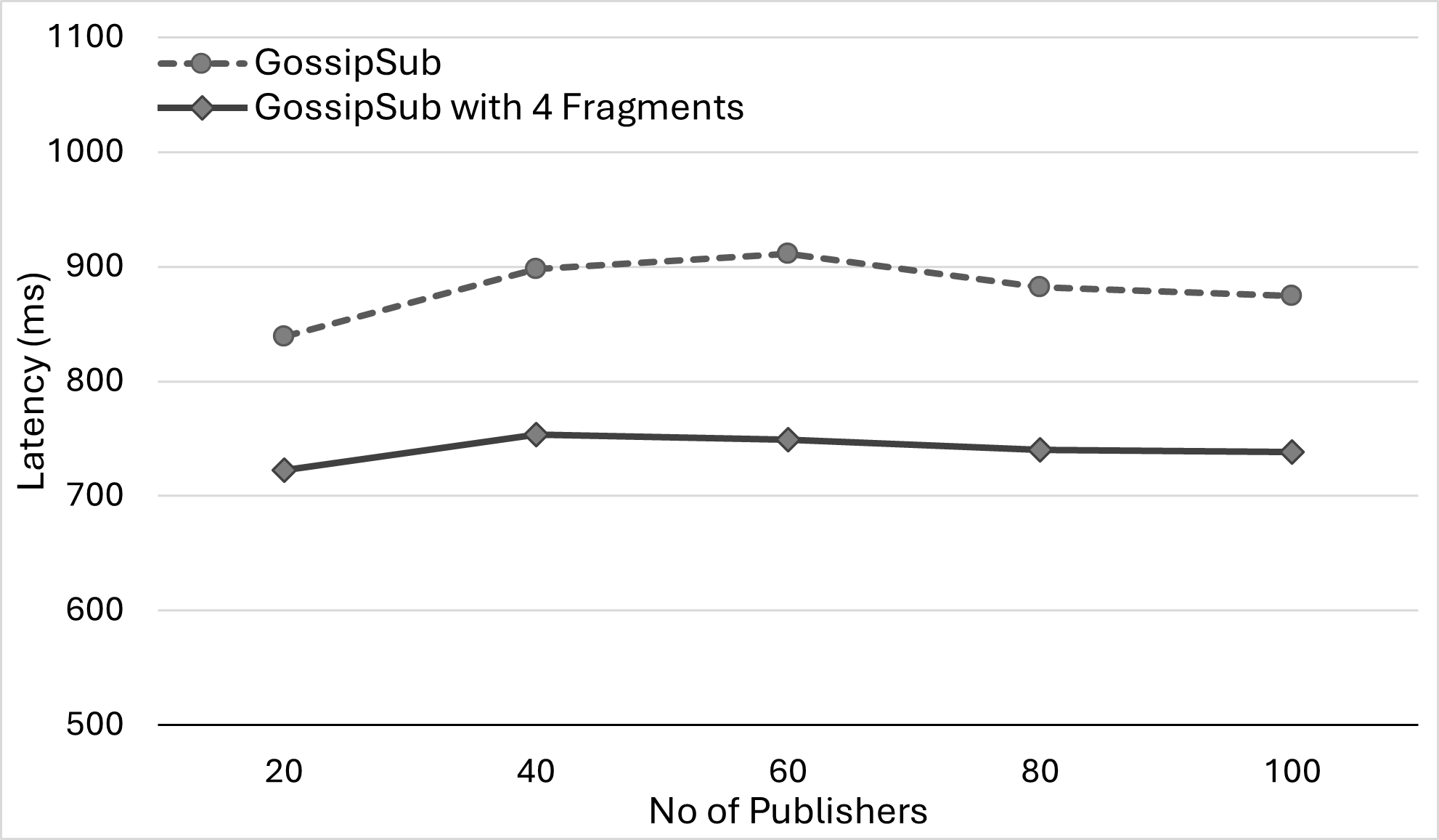}
    }

    \subfigure[Increasing network size (\emph{$B_N$})]{
        \includegraphics[width=0.315\linewidth]{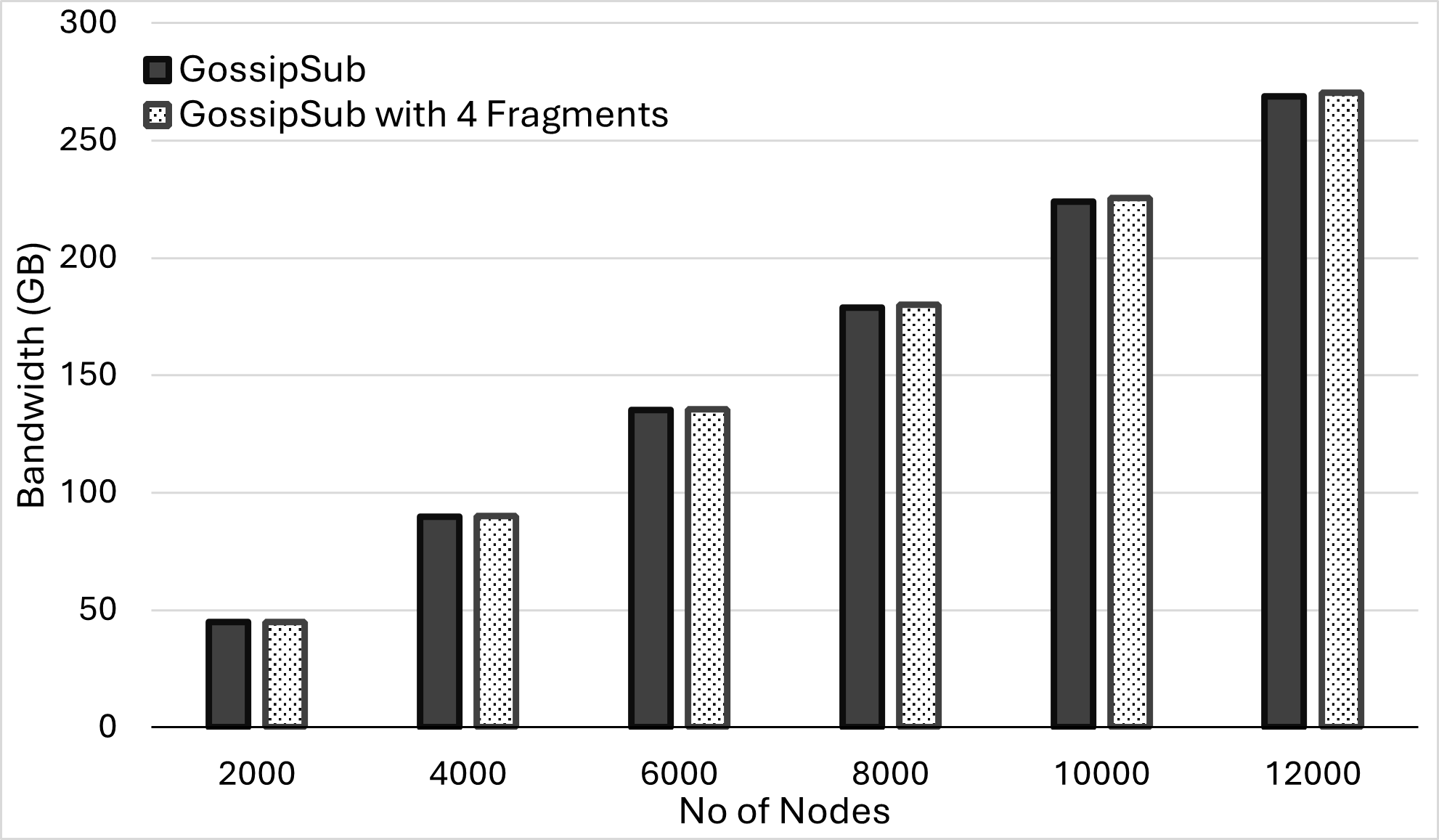}
    }
    \subfigure[Increasing message size (\emph{$B_N$})]{
        \includegraphics[width=0.315\linewidth]{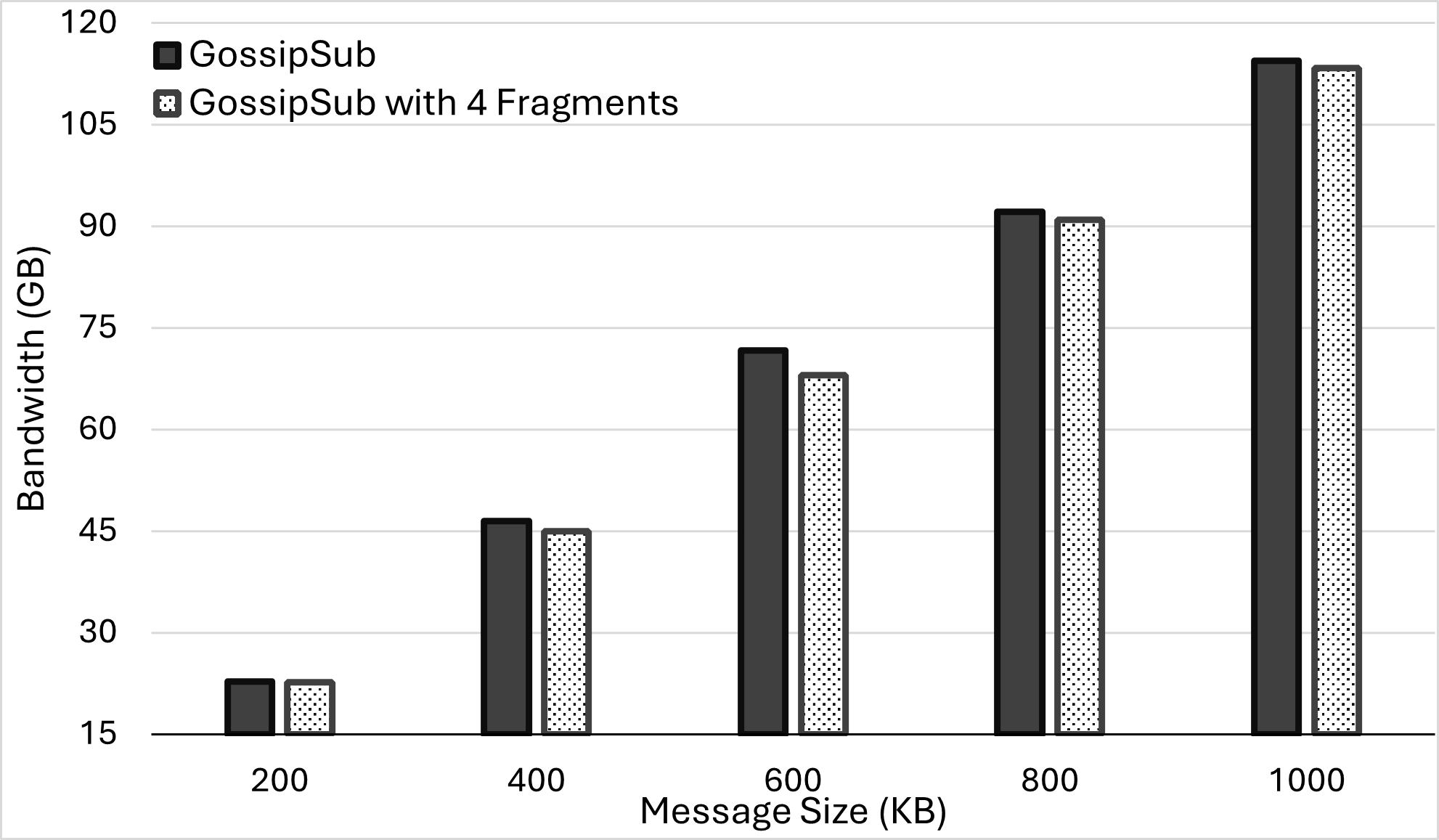}
    }
    \subfigure[Increasing number of publishers (\emph{$B_N$})]{
        \includegraphics[width=0.315\linewidth]{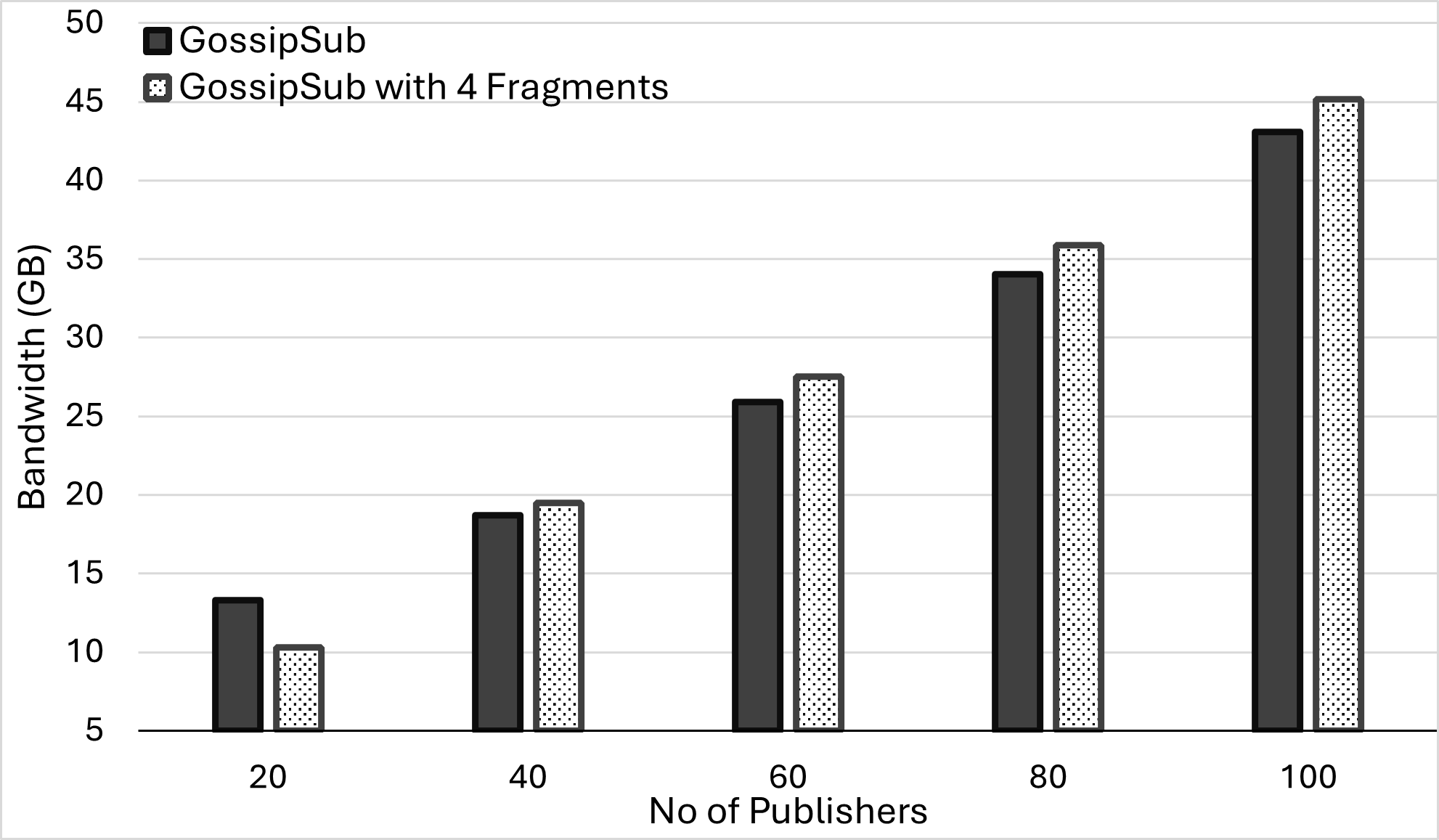}
    }

    \caption{Message fragmentation impact \emph{(using 4 Fragments)}: average Latency (\emph{$L_{cov}^{100}$}) and bandwidth utilization (\emph{$B_N$})}
    \label{fig:gossipsub_fragmentation}
\end{figure*}

\begin{table}[htb]
\centering
\caption{$L_{200-1000KB}^{1000}$ Based on Changing Link Latency}
\begin{tabular}{ccccccc}

\toprule
\multicolumn{1}{c}{} & \multicolumn{3}{c}{GossipSub + IDONTWANT} & \multicolumn{3}{c}{Stagger (Sequential Send)} \\
\cmidrule(lr){2-4} \cmidrule(lr){5-7}
\multicolumn{1}{c}{Message} & \multicolumn{3}{c}{Link Latency} & \multicolumn{3}{c}{Link Latency} \\

\multicolumn{1}{c}{Size(KB)}& \multicolumn{1}{c}{25ms} & \multicolumn{1}{c}{50ms} & \multicolumn{1}{c}{100ms} & \multicolumn{1}{c}{25ms} & \multicolumn{1}{c}{50ms} & \multicolumn{1}{c}{100ms} \\

\hline
200 & 795 & 906 & 1177 & 982 & 1358 & 2037\\
\hline
400 & 2411 & 2523 & 2752 & 2322 & 2845 & 3633\\
\hline
600 & 3813 & 3924 & 4331 & 3309 & 3947 & 5054\\
\hline
800 & 5299 & 5522 & 6051 & 4402 & 5308 & 6569\\
\hline
1000 & 6342 & 6649 & 7008 & 5193 & 6136 & 7393\\
\hline

\end{tabular}
\begin{tablenotes}
    \item[1] We compute $L_{cov}^{100}$ after sending 15 warmup messages. All messages are published by the same publisher. We change link latency to 25, 50, and 100 milliseconds, and both protocols make use of IDONTWANT messages
\end{tablenotes}
\label{tab:warmup_latency}
\end{table}

Message-staggering maximizes the benefits of IDONTWANT messages, as demonstrated in Fig. \ref{fig:stagger_idontwant}(d)-(f). This reduction in bandwidth utilization is attributed to the sequential relaying of messages, which allows most peers to receive and process IDONTWANT notifications from their successors in a timely manner. As a result, using staggering in conjunction with IDONTWANT messages can reduce bandwidth usage by up to 60\% when messages are sent sequentially and up to 38\% when messages are sent to three peers simultaneously. However, malicious senders or receivers can exploit message-staggering to slow down message propagation. Therefore, staggering should be paired with appropriate safety measures, such as peer scoring and timeouts.

\begin{figure*}[!t]
    \centering
    \subfigure[Increasing network size (\emph{$L_{200KB}^{2000-12000}$})]{
        \includegraphics[width=0.315\textwidth]{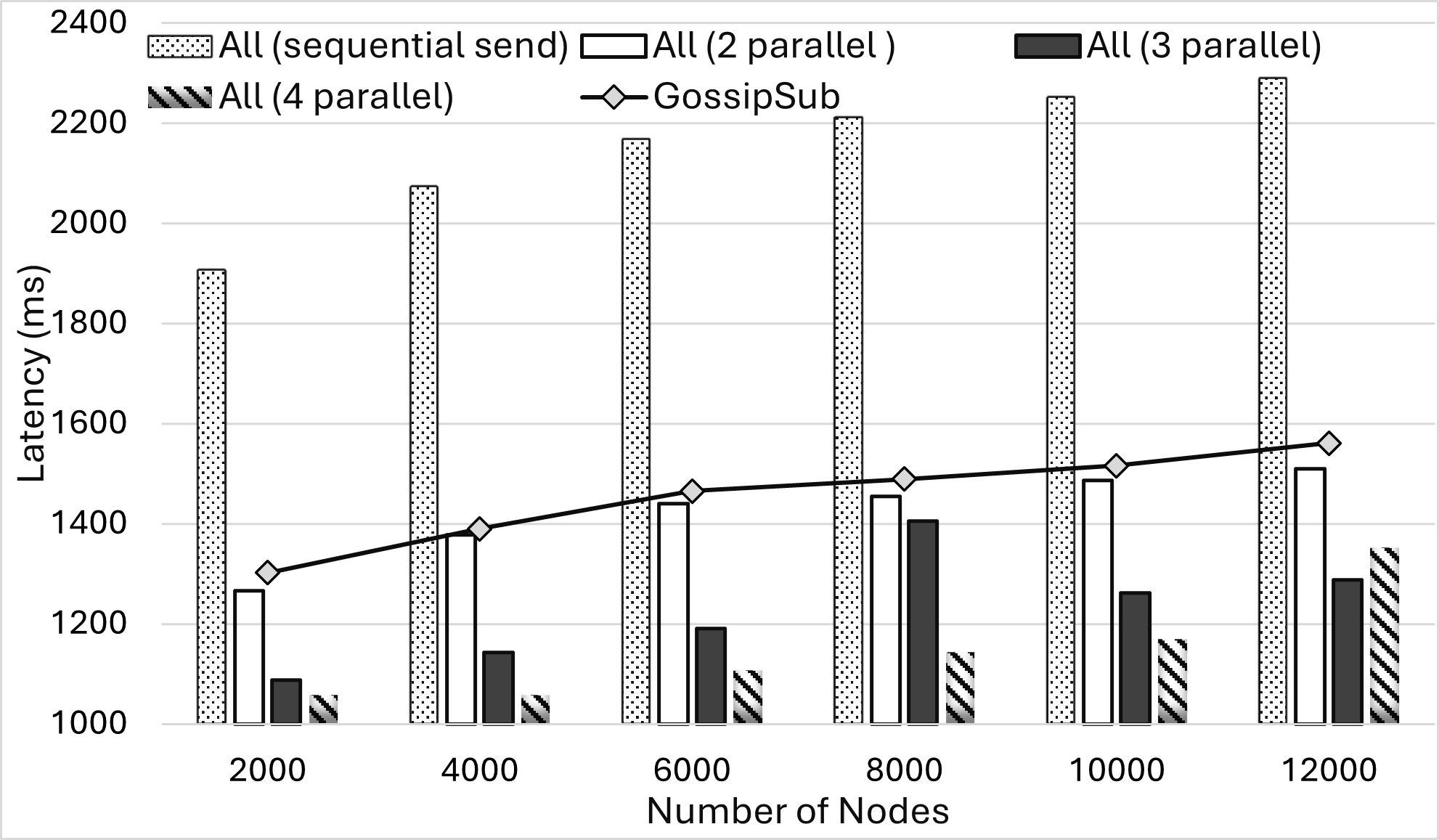}
    }
    \subfigure[Increasing message size (\emph{$L_{200-1000KB}^{1000}$})]{
        \includegraphics[width=0.315\linewidth]{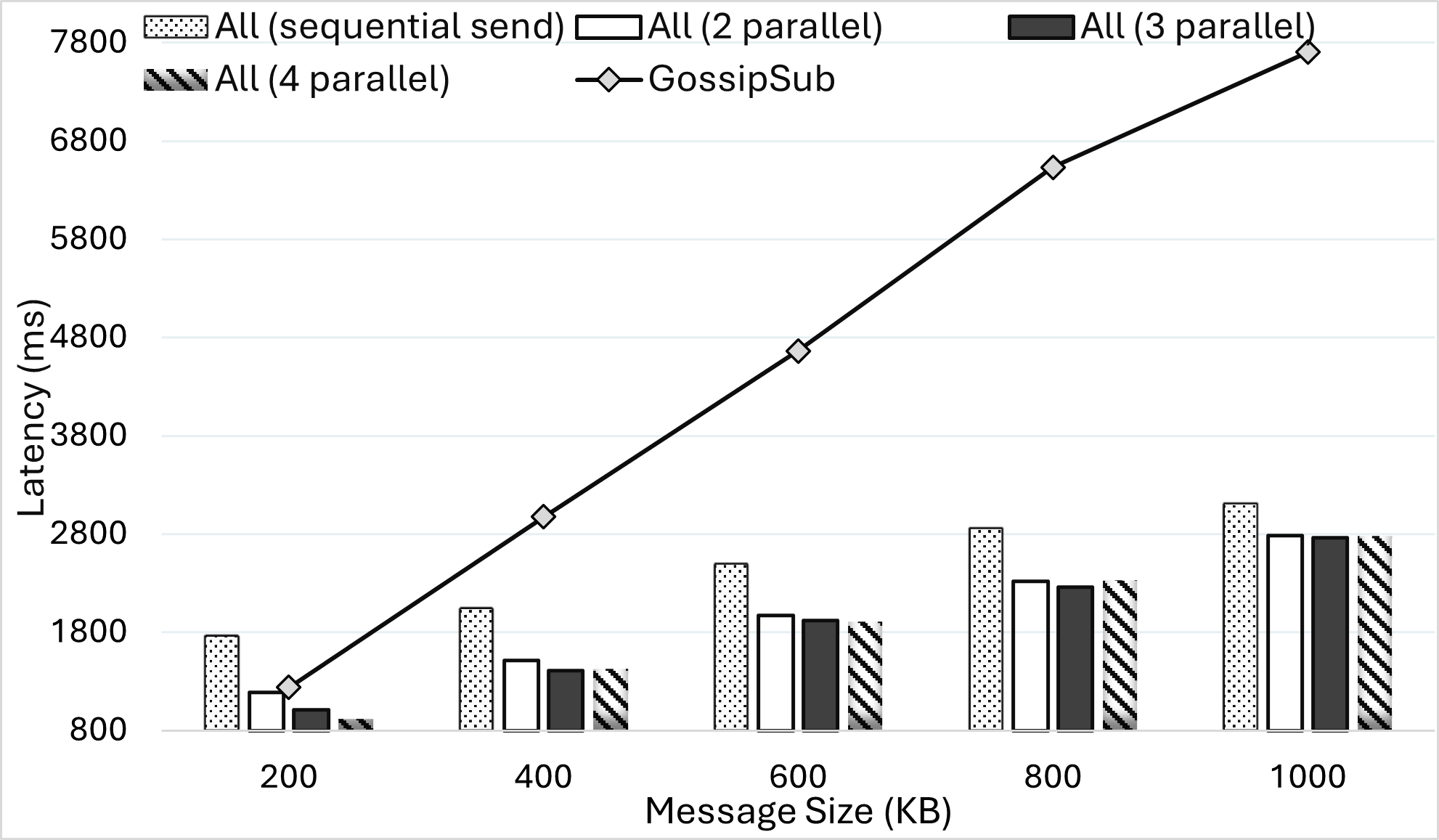}
    }
    \subfigure[Increasing number of publishers (\emph{$L_{50KB}^{1000}$})]{
        \includegraphics[width=0.315\linewidth]{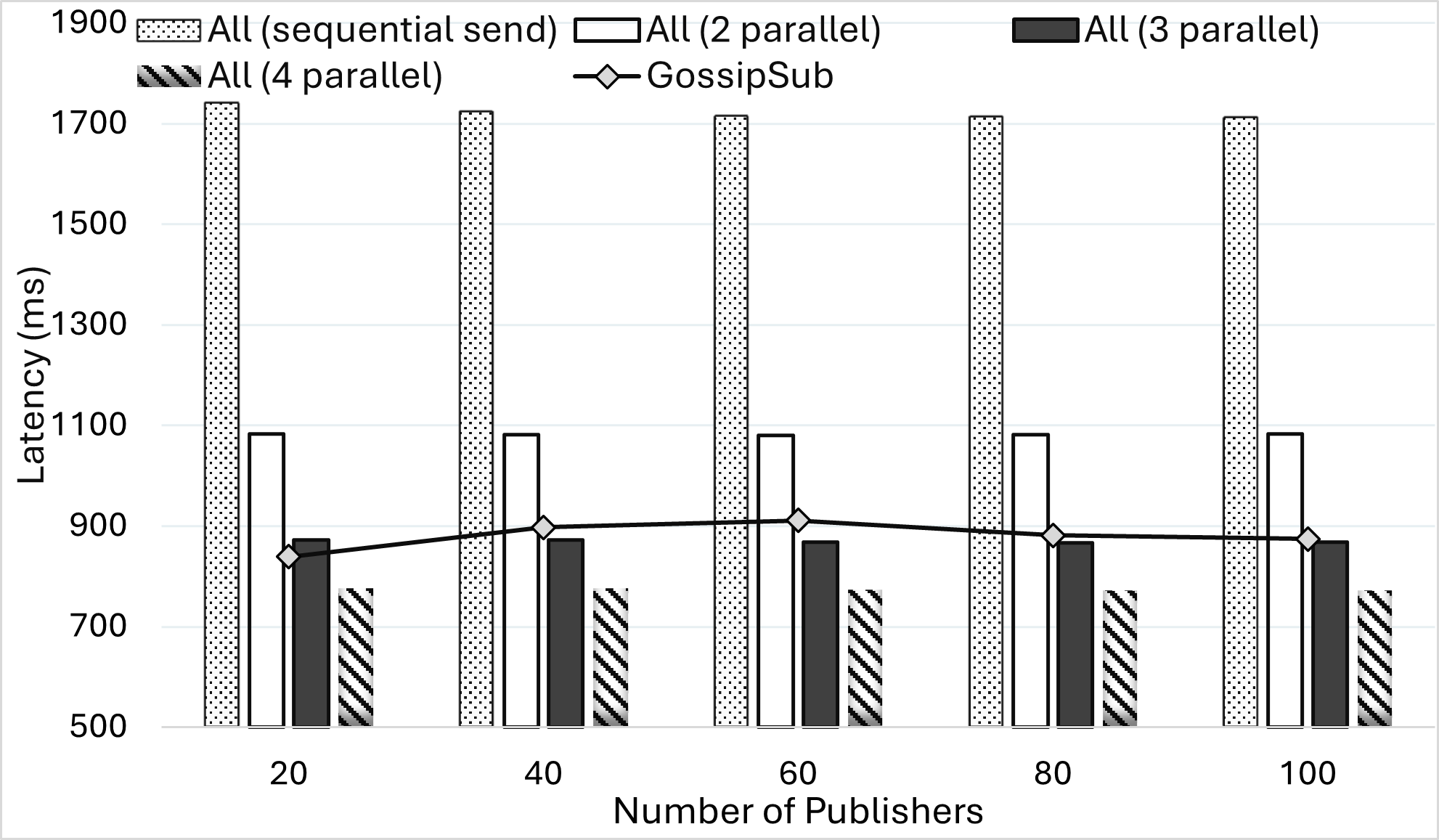}
    }

    \subfigure[Increasing network size (\emph{$B_N$})]{
        \includegraphics[width=0.315\linewidth]{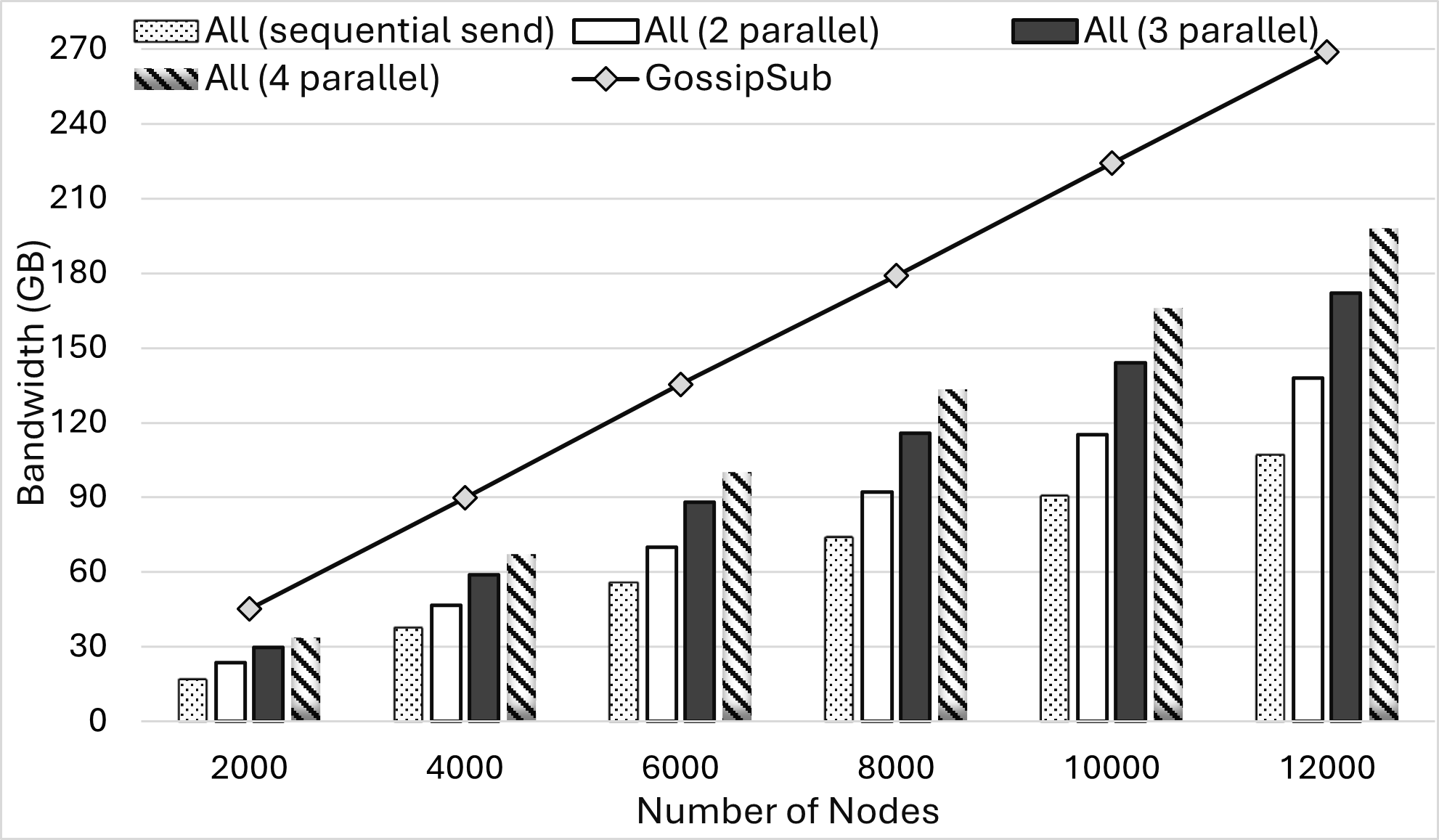}
    }
    \subfigure[Increasing message size (\emph{$B_N$})]{
        \includegraphics[width=0.315\linewidth]{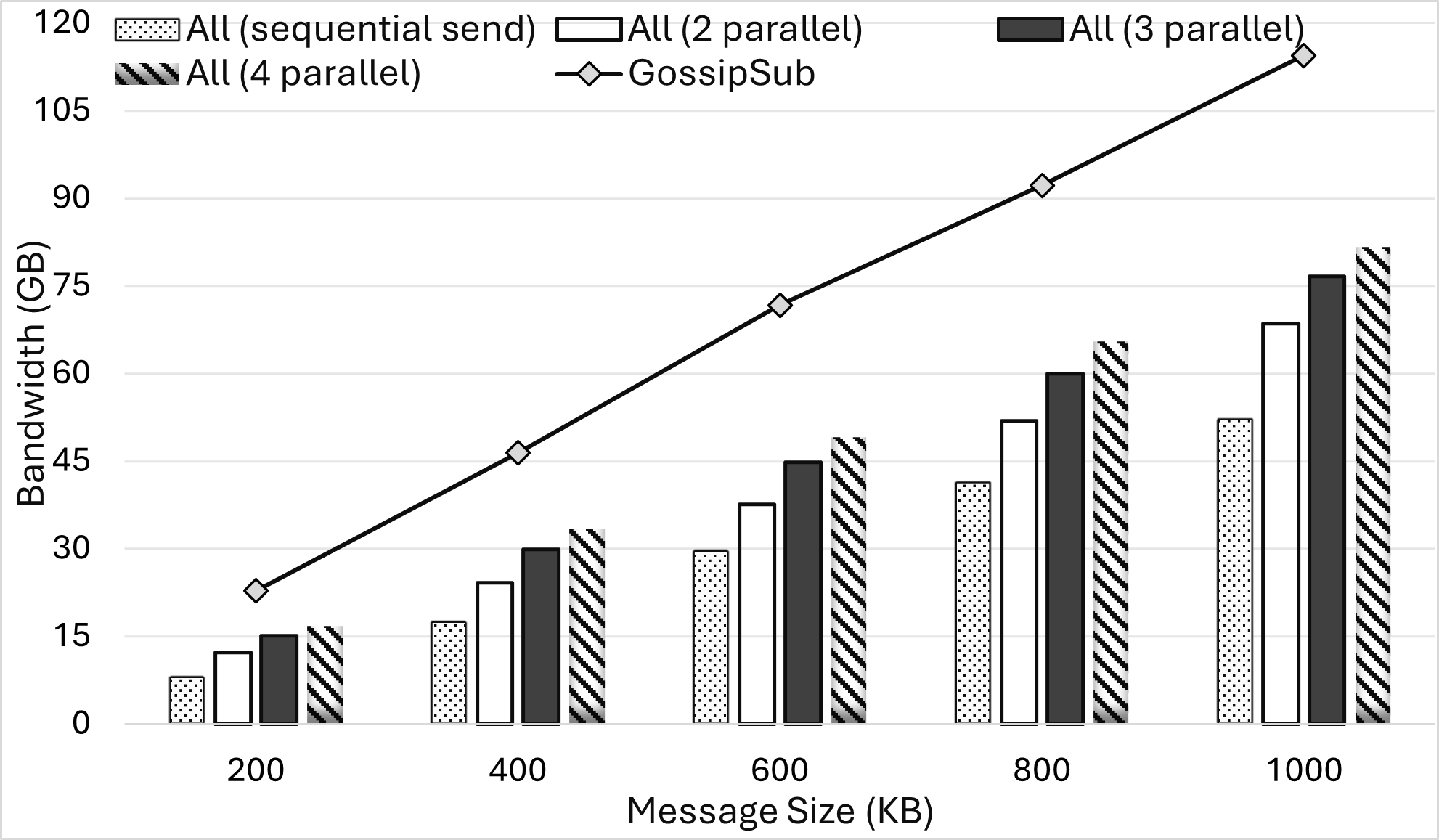}
    }
    \subfigure[Increasing number of publishers (\emph{$B_N$})]{
        \includegraphics[width=0.315\linewidth]{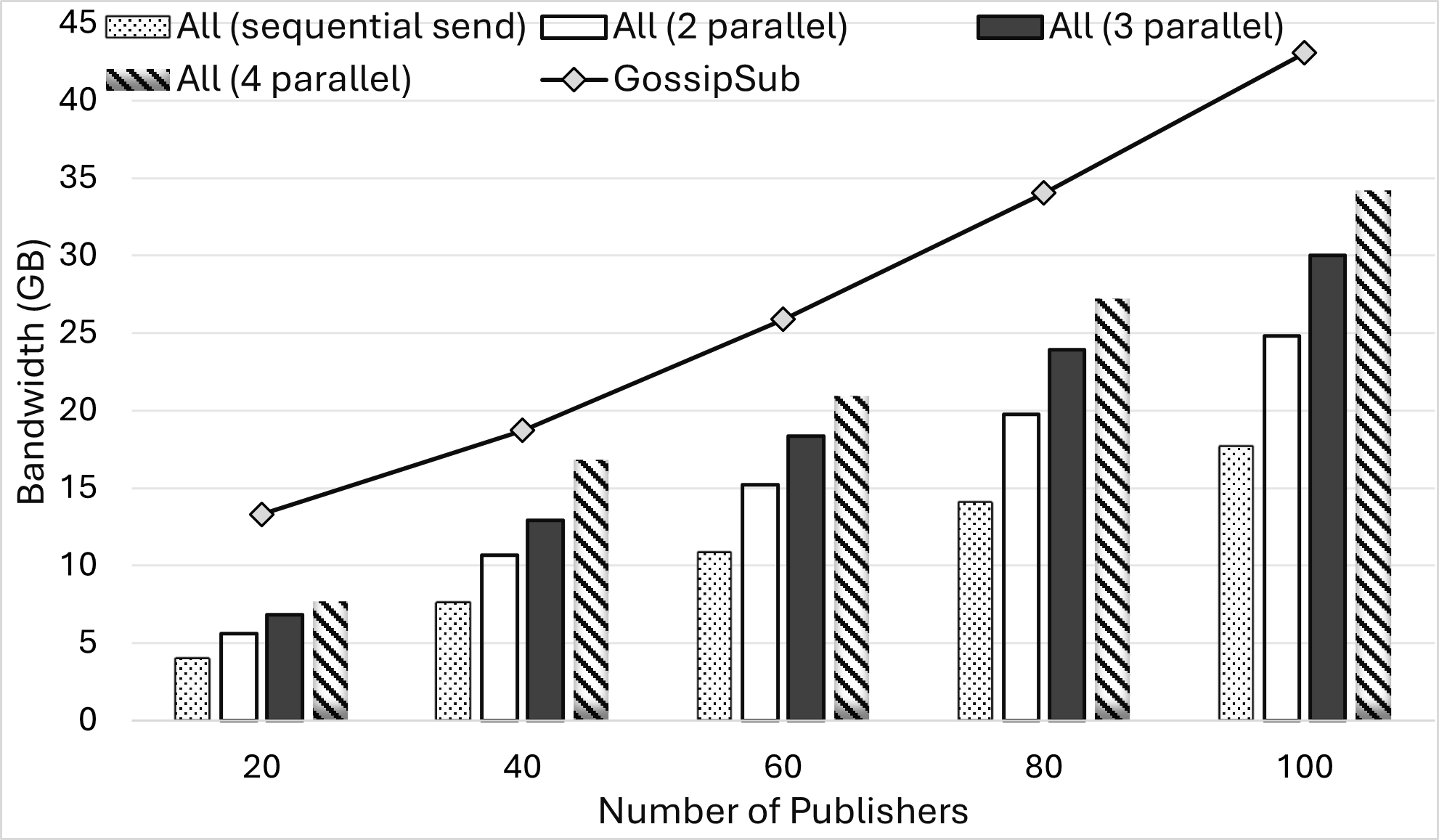}
    }

    \caption{Putting all together \emph{(message-staggering + IDONTWANT + fragmentation)}: average Latency (\emph{$L_{cov}^{100}$}) and bandwidth utilization (\emph{$B_N$})}
    \label{fig:all_together}
\end{figure*}

\begin{table*}[htb]
\centering
\caption{Expanding Coverage Latency ($L_{cov}^{15,85}$) and Latency Deviation $\delta_{L}$}
\begin{tabular}{>{\centering\arraybackslash}p{1.1cm}>{\centering\arraybackslash}m{0.64cm}>{\centering\arraybackslash}m{0.54cm}>{\centering\arraybackslash}m{0.54cm}>{\centering\arraybackslash}m{0.54cm}>{\centering\arraybackslash}m{0.54cm}>{\centering\arraybackslash}m{0.54cm}>{\centering\arraybackslash}m{0.54cm}>{\centering\arraybackslash}m{0.54cm}>{\centering\arraybackslash}m{0.54cm}>{\centering\arraybackslash}m{0.54cm}>{\centering\arraybackslash}m{0.54cm}>{\centering\arraybackslash}m{0.54cm}>{\centering\arraybackslash}m{0.54cm}>{\centering\arraybackslash}m{0.54cm}>{\centering\arraybackslash}m{0.54cm}>{\centering\arraybackslash}m{0.54cm}}

\toprule
\multicolumn{2}{c}{Experiments} & \multicolumn{3}{c}{GossipSub} & \multicolumn{3}{c}{4 Fragments} & \multicolumn{3}{c}{All (Sequential Send)} & \multicolumn{3}{c}{All (2 Parallel)} & \multicolumn{3}{c}{All (4 Parallel)}\\
\cmidrule(lr){3-5} \cmidrule(lr){6-8} \cmidrule(lr){9-11} \cmidrule(lr){12-14} \cmidrule(lr){15-17}
\multicolumn{2}{c}{\& Parameters}& \multicolumn{1}{c}{$L_{cov}^{15}$}\tnote{1} & \multicolumn{1}{c}{$L_{cov}^{85}$} & \multicolumn{1}{c}{$\delta_{L}$} & \multicolumn{1}{c}{$L_{cov}^{15}$} & \multicolumn{1}{c}{$L_{cov}^{85}$} & \multicolumn{1}{c}{$\delta_{L}$} & \multicolumn{1}{c}{$L_{cov}^{15}$} & \multicolumn{1}{c}{$L_{cov}^{85}$} & \multicolumn{1}{c}{$\delta_{L}$} & \multicolumn{1}{c}{$L_{cov}^{15}$} & \multicolumn{1}{c}{$L_{cov}^{85}$} & \multicolumn{1}{c}{$\delta_{L}$} & \multicolumn{1}{c}{$L_{cov}^{15}$} & \multicolumn{1}{c}{$L_{cov}^{85}$} & \multicolumn{1}{c}{$\delta_{L}$}\\
\midrule
\multirow{6}{1.1cm}{\centering Scenario1\\[2ex]Number of\\Nodes} & 2000 & 6 & 9 & 12 & 6 & 8 & 42 & 14 & 17 & 26 & 9 & 10 & 23 & 6 & 8 & 35 \\
& 4000 & 7 & 10 & 28 & 7 & 9 & 31 & 16 & 18 & 31 & 10 & 11 & 45 & 7 & 8 & 20 \\
& 6000 & 7 & 10 & 40 & 7 & 9 & 35 & 17 & 19 & 22 & 8 & 9 & 37 & 7 & 9 & 20 \\
& 8000 & 7 & 11 & 34 & 7 & 9 & 30 & 17 & 19 & 50 & 8 & 10 & 49 & 7 & 9 & 37 \\
& 10000 & 8 & 11 & 23 & 7 & 9 & 53 & 18 & 20 & 46 & 9 & 10 & 39 & 8 & 9 & 61 \\
& 12000 & 8 & 11 & 65 & 7 & 9 & 32 & 18 & 20 & 67 & 9 & 10 & 49 & 8 & 9 & 559 \\
\midrule
\multirow{5}{1.1cm}{\centering Scenario2\\[2ex]Message\\Size (KB)} & 200 & 5 & 9 & 37 & 6 & 8 & 61 & 13 & 15 & 33 & 6 & 8 & 45 & 6 & 7 & 26 \\
&400 & 16 & 23 & 138 & 11 & 16 & 148 & 14 & 17 & 106 & 8 & 11 & 60 & 8 & 11 & 60 \\
& 600 & 26 & 37 & 261 & 16 & 22 & 168 & 16 & 19 & 101 & 11 & 15 & 102 & 11 & 15 & 57 \\
& 800 & 36 & 54 & 614 & 18 & 24 & 167 & 18 & 22 & 89 & 13 & 18 & 112 & 13 & 18 & 122 \\
& 1000 & 41 & 65 & 1018 & 20 & 28 & 173 & 20 & 25 & 160 & 15 & 21 & 229 & 15 & 23 & 234 \\
\midrule
\multirow{5}{1.1cm}{\centering Scenario3\\[2ex]Number of\\Publishers} & 20 & 4 & 6 & 107 & 3 & 5 & 54 & 15 & 17 & 24 & 5 & 7 & 27 & 5 & 6 & 12 \\
& 40 & 4 & 7 & 104 & 4 & 6 & 58 & 13 & 15 & 37 & 5 & 7 & 26 & 5 & 6 & 12 \\
& 60 & 4 & 6 & 99 & 4 & 5 & 54 & 12 & 15 & 46 & 5 & 7 & 30 & 5 & 6 & 17 \\
& 80 & 4 & 6 & 107 & 4 & 5 & 52 & 12 & 15 & 49 & 5 & 7 & 32 & 5 & 6 & 19 \\
& 100 & 4 & 6 & 99 & 4 & 5 & 49 & 12 & 15 & 47 & 5 & 7 & 30 & 5 & 6 & 17 \\
\bottomrule
\end{tabular}
\begin{tablenotes}
    \item[1] The reported values for $L_{cov}^{15}$ and $L_{cov}^{85}$ represent counts over 100-millisecond intervals (they should be multiplied by 100 ms to obtain time estimates).
    \item[2] Identifying peers who are currently receiving messages is challenging, so such peers are not included in the estimates for $L_{cov}^{15}$ and $L_{cov}^{85}$. 
\end{tablenotes}
\label{tab:fullmessage}
\end{table*}

Message fragmentation effectively reduces the store-and-forward delay by partitioning a large message into smaller chunks. Fig. \ref{fig:gossipsub_fragmentation}(b) illustrates that partitioning a 1MB message into four chunks achieves up to 55\% reduction in latency while maintaining a similar bandwidth utilization. Fig. \ref{fig:gossipsub_fragmentation} depicts that message fragmentation effectively reduces latency under varying network conditions. However, the benefits of fragmentation become more apparent when applied to larger messages or in bigger networks, as partitioning a larger message or forwarding fragments through a network with a bigger diameter can help minimize store-and-forward delay.

Combining message-staggering, IDONTWANT technique, and message fragmentation produces very encouraging results, as shown in Fig. \ref{fig:all_together}. Notably, the 'All with 3 parallel sends' approach reduces 1MB message transmission latency ($L_{1MB}^{1000}$) by more than 64\%. Fragmentation alone results in 56\% reduction in ($L_{1MB}^{1000}$), as illustrated in Fig. \ref{fig:gossipsub_fragmentation}(b). In contrast, $L_{1MB}^{1000}$ reported by message-staggering (with 3 parallel sends) is rather higher than that of GossipSub, as shown in Fig. \ref{fig:stagger_idontwant}(b). This implies that the message-staggering approach achieves better $L_{cov}^{100}$ in conjunction with message fragmentation and IDONTWANT techniques. At the same time, it maintains its bandwidth-saving characteristics. This is because peers transmit messages sequentially in staggered sending and remain idle until they receive a new message. On the other hand, fragmentation allows multiple chunks to be propagated simultaneously. This allows staggered message sending and rotation between the messages in the outgoing message queue. Creating this rotational sending mechanism provides numerous benefits over the standard GossipSub operation: 1) Multiple fragments propagate together in the network, creating a situation where different chunks are relayed early by different peers in the mesh. 2) There is an increased chance of receiving and processing IDONTWANT messages. 3) A reduction in store-and-forward delay is achieved. 4) Distant peers in the message propagation path can use IWANT requests to fetch missing fragments. Downloading message fragments requires a relatively shorter time.

The expanding coverage latency $L_{cov}^{15, 85}$ in Table \ref{tab:fullmessage} provides insight into the initial and final phases of message propagation. Additionally, latency deviation $\delta_{L}$ is also reported to highlight fluctuations in $L_{cov}^{100}$. It is important to note that a significant amount of time is consumed in reaching the first 15\% of peers. The all-in-one approach not only results in a faster spread during the earlier stages of message propagation but also tends to maintain consistent message propagation times.

\section{Conclusion and Future Work} \label{S4}
We investigated the challenges of handling large messages in GossipSub and experimented with various possible optimizations to reduce message dissemination time and bandwidth utilization. We considered several factors, including congestion avoidance mechanisms, redundant transmissions, message forwarding strategies, increased message transmission times, and store-and-forward delays, to gain a comprehensive understanding of the problem. Larger message sizes lead to longer transmission times that accumulate over multi-hop paths. At the same time, GossipSub features like IWANT requests do not account for message size. TCP congestion avoidance algorithms also increase message transmission times over newly established (cold) connections. Coupling these constraints with the probabilistic message-forwarding nature of GossipSub complicates the situation by utilizing a considerable share of available bandwidth on redundant transmissions.

Using IDONTWANT messages helps reduce many of these redundant transmissions. However, it does not have much impact on message dissemination latency. Message-staggering further minimizes bandwidth utilization by maximizing the performance benefits of IDONTWANT messages. However, sub-optimal peer prioritization, numerous IWANT requests, and congestion avoidance mechanisms compromise the expected improvements in message dissemination latency. On a positive note, message fragmentation considerably reduces message dissemination latency by lowering store-and-forward delays. However, it does not have much impact on bandwidth utilization. Combining these approaches lowers both bandwidth utilization and message dissemination latency. This is because fragmentation reduces store-and-forward delay and propagates multiple fragments in the network. Using message-staggering with rotational sending and IDONTWANT messages lowers bandwidth utilization and achieves quicker message distribution. However, fragmentation and message-staggering must be used with caution. Fragmentation may allow non-conforming peers to stop/delay relaying some fragments. Similarly, message-staggering may enable malicious peers to deliberately slow down message propagation.

The performance evaluations are conducted in a simplified simulation environment to reduce the impact of variables like dissimilar path characteristics. Future work will explore the proposed schemes under realistic conditions, including latency/bandwidth variations, and in the presence of adversaries.

\section*{Acknowledgment}
The authors thank Anton Nashatyrev for his engagement and valuable feedback. The authors also acknowledge that this work builds upon the foundation laid in https://hackmd.io/X1DoBHtYTtuGqYg0qK4zJw and https://hackmd.io/@nashatyrev/B18wdnNDh

\bibliographystyle{IEEEtran}
\bibliography{mybib}

\end{document}